\documentclass[10pt,journal]{IEEEtran}%lettersize,journal / 12pt,onecolumn,
\usepackage{amsmath,amssymb,amsfonts,amsthm}
\usepackage{array}
\usepackage[caption=false,font=footnotesize,labelfont=rm,textfont=rm]{subfig}
\usepackage{textcomp}
\usepackage{stfloats}
\usepackage{url}
\usepackage{cases}

\usepackage{multirow}
\usepackage{xcolor}
\usepackage{verbatim}
\usepackage{graphicx}

\usepackage{cite}
\usepackage{threeparttable}
\usepackage{booktabs}
\usepackage{pifont}

\usepackage[ruled]{algorithm2e} %,vlined

\SetCommentSty{mycommfont}

\def\BibTeX{{\rm B\kern-.05em{\sc i\kern-.025em b}\kern-.08em
    T\kern-.1667em\lower.7ex\hbox{E}\kern-.125emX}}
\usepackage{balance}

\begin{document}
\bstctlcite{setting}
\setlength{\textfloatsep}{5pt}
\setlength{\floatsep}{5pt}
\setlength{\abovedisplayskip}{6pt}
\setlength{\belowdisplayskip}{6pt}

%\title{Intelligent Deployment of Urban Macro Base Stations via Wireless Network Digital Twin: \\A Geographic Data-Informed Approach}

\title{Intelligent Base Station Deployment in Urban Wireless Networks: A Geographic Data-Informed Digital Twin Approach}

\author{Zhenyu~Tao,~\IEEEmembership{Graduate Student Member,~IEEE},
Yuxuan~Li,~\IEEEmembership{Graduate Student Member,~IEEE},
  Wei~Xu,~\IEEEmembership{Fellow,~IEEE},
  Yongming~Huang,~\IEEEmembership{Fellow,~IEEE},
  and Xiaohu~You,~\IEEEmembership{Fellow,~IEEE}% <-this % stops a space and~

  %\thanks{\textit{(Corresponding author: Wei Xu,Xiaohu You.)}}
  %\thanks{(\textit{Corresponding author: Xiaohu You.})}
  \thanks{Z. Tao, Y. Li, W. Xu, Y. Huang, and X. You are with the National Mobile Communications Research Lab, Southeast University, Nanjing 210096, China, and also with the Pervasive Communication Research Center, Purple Mountain Laboratories, Nanjing 211111, China (email: \{zhenyu\_tao, yuxuan\_li, wxu, huangym, xhyu\}@seu.edu.cn).} %, \textit{Corresponding authors: Wei Xu, Xiaohu You.
}

% \markboth{IEEE TRANSACTIONS ON , ~Vol.~14, No.~8, August~2021}%
% {Shell \MakeLowercase{\textit{et al.}}: A Sample Article Using IEEEtran.cls for IEEE Journals}

\maketitle

\begin{abstract}
Base station (BS) placement fundamentally determines urban wireless network coverage and capacity. However, its optimization remains challenging because it depends on site-specific propagation and user distribution information that is difficult to obtain before deployment. To address this challenge, we propose a geographic data-informed wireless network digital twin (DT) framework that incorporates a sample-free radio map prediction model with a hybrid input representation to achieve kilometer-scale signal strength estimation in milliseconds, complemented by a diffusion-based trajectory generation model to collectively characterize channel and user distributions. With pretrained DT models, optimization requires only target-area geographic data, without measurements, ray tracing (RT), or real user trajectories.
BS deployment is formulated as a multi-step Markov decision process and optimized using spatial proximal policy optimization within the DT environment. A Wasserstein distance-based buffer is introduced to retain high-quality, spatially distinct deployments for multi-start local search, combining policy exploration with fine-grained deployment refinement.
Experiments demonstrate sample-free radio map accuracy approaching prediction with 100 sparse received signal strength samples. In held-out test areas, the intelligent BS deployment framework outperforms traditional and learning-based baselines, achieving approximately 89.2\%--97.0\% of the near-optimal RT-based performance while reducing optimization time from 31.5 hours to approximately four minutes per area. Cross-city transfer without DT model fine-tuning also achieves 87.4\%--98.6\% of the near-optimal performance in less than three minutes per area.
\end{abstract}


\section{Introduction}\label{sec:intro}
\IEEEPARstart{T}{he} vision of 6G aims to deliver extreme connectivity, ultra-broadband connectivity, and deep integration of artificial intelligence (AI)~\cite{10183795,11381448}.
To realize this, extensive research has focused on post-deployment optimization techniques, such as resource scheduling~\cite{8796358}, beamforming~\cite{9303466}, and user association~\cite{9127161}, to enhance quality of service (QoS).
While effective, these strategies are inherently constrained by the physical placement of base stations (BSs), as deployment locations critically determine achievable network performance. In this context, effective and efficient BS deployment serves as a critical foundation for all subsequent optimization, particularly in dense urban scenarios where complex building landscapes severely affect signal propagation~\cite{borralho_survey_2021}.

Existing BS deployment optimization studies, ranging from traditional heuristic and stochastic geometry methods~\cite{coskun_energy-efficient_2014,zhang_energy_2016} to recent convex optimization~\cite{diamantoulaki_optimal_2025} and deep reinforcement learning (DRL) approaches~\cite{zhu_deep_2025}, typically rely on either statistical channel models for tractable theoretical analysis or on assumed prior knowledge of channel gains.
However, stochastic-model-based methods often result in suboptimal deployment, as physical propagation in complex urban environments seldom conforms to idealized statistical models~\cite{borralho_survey_2021}, while the latter assumptions face a practical limitation, that is, the site-specific information necessary for optimization is inherently unavailable prior to deployment~\cite{lee_autobs_2025}.
Although such information could, in principle, be obtained through extensive on-site measurements, the vast number of candidate locations, coupled with the complexity of multi-BS coordination, results in prohibitive measurement overhead in practice~\cite{8648450}.
%,tao2025tsp

To obtain site-specific channel information in BS deployment optimization without costly measurements, digital twin (DT) technology offers a promising solution by creating virtual replicas of the wireless environment and establishing radio maps to capture spatial signal characteristics~\cite{10234388,tao2023wireless}. However, current radio map construction approaches present notable obstacles. Ray tracing (RT) can generate high-fidelity radio maps, but even graphics processing unit (GPU)-accelerated implementations such as Sionna~\cite{sionna} require tens of seconds per radio map~\cite{belgiovine_better_2025}. This computational cost forms a critical bottleneck for large-scale iterative optimization that requires evaluation of numerous candidate configurations~\cite{su_jointly_2026}. While neural network (NN)-based radio map estimation methods have gained increasing attention as a faster alternative, they usually face distinct limitations for BS deployment planning. Specifically, sample-based methods rely on sparse on-site measurements from deployed BSs~\cite{li_geo2sigmap_2024,lin_geo2commap_2025,chen_high-efficiency_2025}, leading to a circular dependency with the deployment process itself, while sample-free methods often adopt simplified urban assumptions, such as uniform building heights and homogeneous building material properties~\cite{levie_radiounet_2021, lee2023pmnet,lee_scalable_2024,jaensch_radio_2025,11174709}, that are inadequate for representing the propagation environment of macro BSs covering kilometer-scale urban areas.

Evaluating network capacity, i.e., the aggregate achievable throughput, for BS deployment further requires knowledge of user spatial distribution, which is tightly coupled with channel conditions and jointly determines link quality, cell load, and bandwidth allocation within the wireless network~\cite{7070658}. Despite this tight coupling, most existing BS deployment strategies model user distribution separately from the wireless channel, typically by assuming either uniform user distributions~\cite{lee_autobs_2025} or stochastic mobility patterns~\cite{zhang_optimal_2021,lyu_spatial_2024} that fail to capture the constraints imposed by urban street topology. As a result, these simplified assumptions can lead to suboptimal deployment decisions that do not reflect actual traffic demand, especially during rush hours when users cluster along road networks and generate severe load imbalance across the network.

To address these challenges, in this paper, we exploit openly available urban geographic data from OpenStreetMap~\cite{haklay2008openstreetmap} to construct the wireless network DT and achieve intelligent BS deployment. The DT models are trained offline using ray-tracing and trajectory data, enabling deployment optimization in the target area to rely solely on geographic inputs. The contributions are summarized as follows:
\begin{itemize}
  \item We propose a geographic data-informed wireless network DT framework for urban macro BS deployment that jointly accounts for radio propagation and spatial user demand. The DT combines building-based radio map prediction with street-map-conditioned trajectory generation to estimate channel conditions and user distributions, and incorporates inter-BS interference and load-dependent bandwidth sharing into a unified coverage-capacity objective. It enables demand-aware deployment optimization without target-area on-site measurements, real user trajectories, or RT during optimization.
  \item To characterize complex radio propagation in kilometer-scale urban scenarios, we develop a sample-free radio map prediction model with a hybrid input representation. By jointly encoding 3D building heights, BS locations and antenna heights, and material-specific electromagnetic (EM) properties, the representation captures geographic information essential to height-dependent propagation and multiple reflections in macrocell scenarios. We further construct and publicly release a corresponding RT dataset\footnote{\url{https://zenodo.org/records/19202802}.}. The radio map prediction model supports prediction across heterogeneous urban environments and millisecond-level inference for deployment evaluation.
  \item We develop an intelligent BS deployment algorithm that combines policy exploration, deployment diversity preservation, and local refinement within the DT. The deployment task is formulated as a multi-step Markov decision process (MDP), and a proximal policy optimization (PPO) agent uses a shared spatial encoder to process large urban states. A Wasserstein distance-based buffer retains high-quality, spatially distinct deployments for multi-start local search (LS), enabling refinement across multiple promising regions of the combinatorial deployment space.
  \item We comprehensively validate the framework using real geographic data, authentic user trajectories, and RT results. Radio map experiments demonstrate the benefits of the hybrid representation, achieving accuracy close to prediction with 100 sparse received signal strength (RSS) samples. Experiments in held-out test areas establish consistent deployment gains over traditional and DRL baselines, while component-wise ablations verify the benefits of spatial policy learning, LS refinement, and the deployment buffer. The framework approaches the performance of the near-optimal reference while reducing optimization time from over 30 hours to less than 5 minutes. Furthermore, a cross-city transfer experiment demonstrates effective deployment optimization using only target-city geographic data, without DT model fine-tuning.
\end{itemize}

\section{Related Work}\label{sec:related}

\subsection{Base Station Deployment Optimization}
BS deployment has been extensively studied over the past decades, with existing methods broadly categorized into model-driven analytical approaches and data-driven learning-based approaches. Early analytical methods employed heuristic algorithms~\cite{coskun_energy-efficient_2014} and stochastic geometry-based analyses~\cite{zhang_energy_2016}, and more principled formulations employed convex optimization techniques with Karush-Kuhn-Tucker (KKT) conditions~\cite{diamantoulaki_optimal_2025}. More recently, the rise of AI~\cite{10024766} has motivated DRL-based BS placement methods~\cite{zhu_deep_2025} that can adapt to complex deployment scenarios. However, these methods either rely on statistical models~\cite{coskun_energy-efficient_2014,zhang_energy_2016} that are too idealized to represent the site-specific wireless environment, or assume the availability of detailed channel information~\cite{diamantoulaki_optimal_2025,zhu_deep_2025}, which is often difficult to obtain during the planning phase. AutoBS~\cite{lee_autobs_2025} and AutoPlan~\cite{11587682} represent recent efforts to integrate DTs with network planning. AutoBS combines PPO with learned radio map prediction, whereas AutoPlan employs a calibrated digital radio twin with Bayesian optimization. However, both of them rely on simplified capacity models based on signal-to-noise ratio (SNR), neglecting inter-BS interference and user-dependent resource sharing, which motivates a BS deployment optimization method that jointly accounts for propagation, interference, and spatial user demand.

\subsection{Radio Map Construction}
RT is widely regarded as the benchmark for generating high-fidelity deterministic channel models~\cite{11173662}, yet its computational cost~\cite{belgiovine_better_2025} remains prohibitive for large-scale deployment optimization. To accelerate radio map generation, NN-based methods have been developed along two main directions. Sample-based approaches reconstruct radio maps from sparse on-site measurements~\cite{li_geo2sigmap_2024,lin_geo2commap_2025,chen_high-efficiency_2025}, achieving competitive accuracy but creating a circular dependency with deployment planning, as measurement data are inherently unavailable before BSs are deployed. Sample-free approaches predict radio maps directly from building geometry without any measurements~\cite{levie_radiounet_2021,lee2023pmnet,lee_scalable_2024,jaensch_radio_2025,11174709}, thereby eliminating this dependency. However, existing sample-free methods generally rely on simplified urban assumptions, such as limited building height variation and homogeneous EM material properties across surfaces, as summarized in Table~\ref{table:comparison}. While such simplifications may be acceptable for pico or micro BSs covering areas below $256\times256$~$\text{m}^2$, they become inadequate for commercial macro BSs~\cite{huawei_aau5613} that can operate at up to 53~dBm radio-frequency (RF) output power and cover kilometer-scale areas, where large height variations and diverse building materials critically affect propagation. To date, existing sample-free radio map prediction methods remain insufficiently accurate for large-scale urban macro BS deployment optimization.
\begin{table*}[!t]
  \centering
  \caption{Comparison of Radio Map Prediction Models}
  \label{table:comparison}
  \renewcommand{\arraystretch}{0.7} % Improves vertical spacing
  \begin{tabular}{lcccccc}
    \toprule
    \textbf{Model}                                        & \textbf{Methodology} & \textbf{Building Representation} & \textbf{Coverage Area}                 & \textbf{Building Material} & \textbf{BS Height} & \textbf{BS Type} \\
    \midrule
    Geo2SigMap~\cite{li_geo2sigmap_2024}                  & Sample-based         & 2D                          & $512\times 512$~$\text{m}^2$           & Uniform                    & Fixed              & Micro ($40$~dBm)          \\
    Geo2ComMap~\cite{lin_geo2commap_2025}                 & Sample-based         & 2D                          & $512\times 512$~$\text{m}^2$           & Uniform                    & Fixed              & Micro ($35$~dBm)          \\
    Chen \textit{et al.}~\cite{chen_high-efficiency_2025} & Sample-based         & 3D ($6\text{-}38.6$~m)      & $300\times 280$~$\text{m}^2$           & Uniform                    & Flexible           & N.A.              \\
    RadioUNet~\cite{levie_radiounet_2021}                 & Sample-free          & 2D                          & $256\times 256$~$\text{m}^2$           & Uniform                    & Fixed              & Pico ($23$~dBm)          \\
    PMNet~\cite{lee2023pmnet,lee_scalable_2024}           & Sample-free          & 2D                          & $221\times 221$~$\text{m}^2$           & Uniform                    & Fixed              & Pico ($23$~dBm)          \\
    UNetDCN~\cite{jaensch_radio_2025}                     & Sample-free          & 3D ($6\text{-}30$~m)        & $256\times 256$~$\text{m}^2$           & Uniform                    & Flexible           & Pico ($30$~dBm)          \\
    \textbf{Proposed}                                     & \textbf{Sample-free} & \textbf{3D (6-300~m)}       & \textbf{1800$\times$1800~$\text{m}^2$} & \textbf{Heterogeneous}     & \textbf{Flexible}  & \textbf{Macro (53~dBm)}   \\
    \bottomrule
  \end{tabular}
\end{table*}

\subsection{User Mobility Modeling}
User mobility shapes the spatiotemporal distribution of traffic demand and directly affects network performance metrics such as handover frequency and average throughput~\cite{8673556}. However, most BS deployment studies overlook its impact to ensure analytical tractability and reduce computational complexity.
For instance, the work in~\cite{lee_autobs_2025} assumes a uniform user distribution over the coverage area, ignoring the dynamic nature of user movements. Stochastic mobility models are also widely used. The random waypoint model~\cite{Johnson1996} is a prevalent choice for coverage analysis and BS deployment~\cite{zhu_deep_2025,sawalmeh_wireless_2019}, while the Gauss-Markov model~\cite{752157} has been employed to capture temporal correlations of user velocity~\cite{zhang_optimal_2021, lyu_spatial_2024}.
However, these models usually fail to capture the road network constraints that govern real urban mobility. It remains impractical to obtain real trajectory data during the pre-deployment phase due to privacy concerns and high collection costs~\cite{8673556}.

\section{Problem Formulation}\label{sec:problem}

This section formulates the BS deployment task as a multi-objective optimization problem that jointly considers network coverage and system capacity.

\begin{figure}[!t]
\centering
\includegraphics[width=0.9\linewidth]{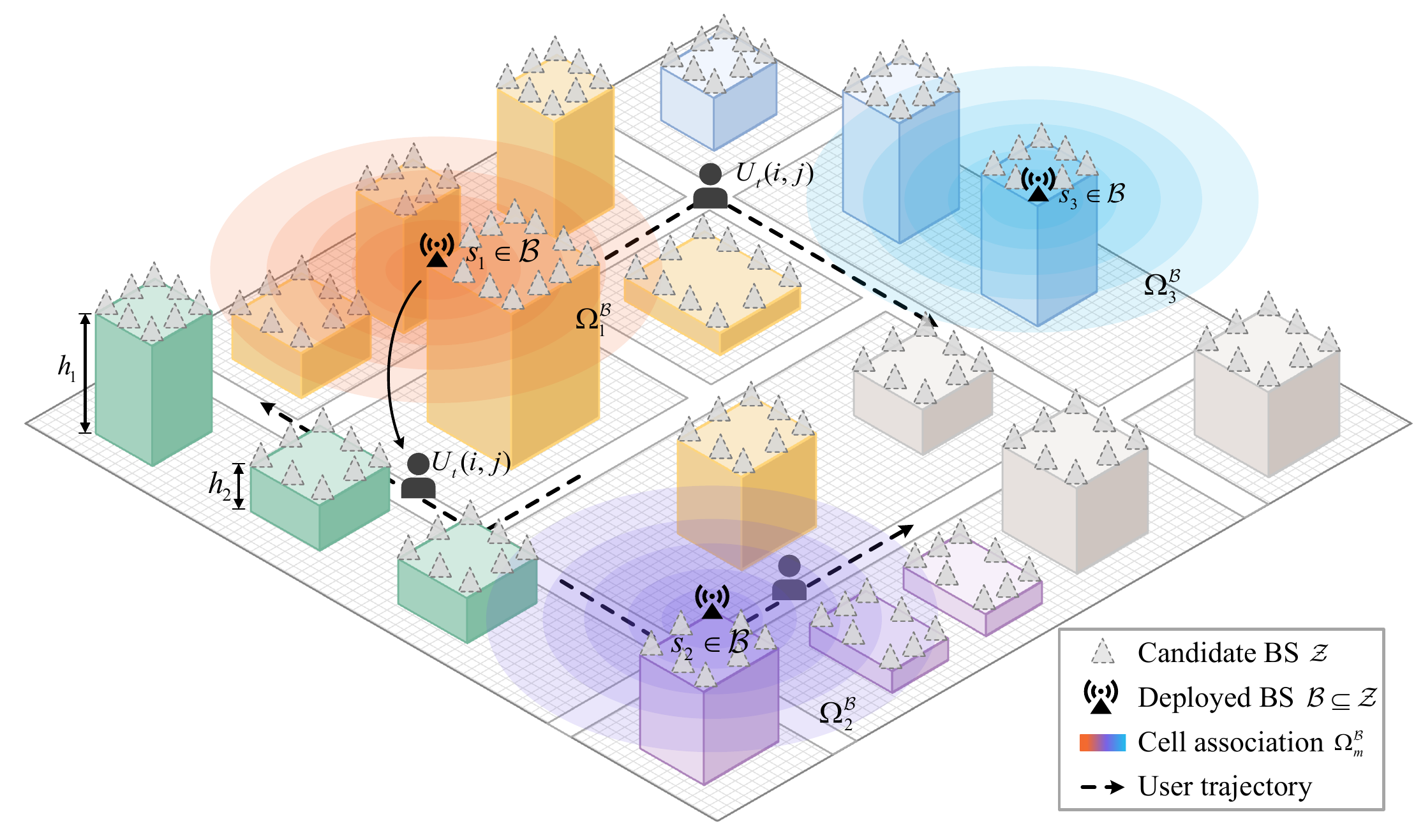}
\caption{BS deployment system model.}
\label{fig:scenario}
\end{figure}

\subsection{System Model}\label{subsec:system_model}
As depicted in Fig.~\ref{fig:scenario}, we consider a square urban area discretized into a grid of $N_{\text{p}} \times N_{\text{p}}$ pixels, where each pixel is indexed by $(i,j)$. The area consists of buildings with varied heights and materials that serve as obstacles to radio propagation. We define the set of outdoor pixels as $\mathcal{O} = \{(i,j) \mid (i,j) \text{ is not occupied by buildings}\}$, encompassing streets and open spaces where users may be located.

A set of $K$ candidate BS locations is defined as $\mathcal{Z} = \{z_1, z_2, \ldots, z_K\}$ at building rooftops, where each candidate $z_k = (x_k, y_k, h_k)$ is characterized by its horizontal coordinates $(x_k, y_k)$ and antenna height $h_k$. The goal is to select a subset $\mathcal{B} \subseteq \mathcal{Z}$ of BSs for deployment, with $|\mathcal{B}|\!=\!M$ and $M\! \ll\! K$, such that the overall network performance is maximized.

Each deployed BS transmits with power budget $P$ over a system bandwidth $B$. The path loss $\text{PL}_m(i,j)$ from BS $m$ to pixel $(i,j)$ captures the aggregate propagation effects, including free-space attenuation, reflection, and diffraction induced by the surrounding building geometry and EM properties.

A total of $N_\text{u}$ users are distributed across the network. The user count map $U_t(i,j)$ denotes the number of users at pixel $(i,j) \in \mathcal{O}$ at time $t$, satisfying $\sum_{i,j} U_t(i,j) = N_\text{u}$. Due to user mobility, the spatial distribution $U_t(i,j)$ can vary dynamically over time.

\subsection{Optimization Target}

\subsubsection{Coverage Objective}\label{subsec:coverage}
Coverage quantifies the fraction of outdoor area receiving adequate signal strength from the deployed BSs. We denote the RSS at pixel $(i,j)$ from BS $m$ as $\psi_{m}(i,j)$, calculated as the difference between the transmit power and the path loss. Considering the max-RSS association principle, the coverage indicator is defined as
\begin{equation}
  \mathcal{I}_{\text{cov}}(i,j,\mathcal{B}|\boldsymbol{\psi}) = \mathbb{I}\left(\max\nolimits_{m \in \mathcal{B}} \psi_{m}(i,j) \geq \psi_{\text{thr}}\right),
  \label{eq:cov_indicator}
\end{equation}
where $\boldsymbol\psi$ denotes the RSS maps, $\mathbb{I}(\cdot)$ denotes the indicator function, and $\psi_{\text{thr}}$ is the minimum sensitivity threshold. The overall network coverage ratio is then evaluated by
\begin{equation}
  R_{\text{cov}}(\mathcal{B}|\boldsymbol{\psi}) = {\sum\nolimits_{(i,j) \in \mathcal{O}} \mathcal{I}_{\text{cov}}(i,j,\mathcal{B}|\boldsymbol{\psi})}/|\mathcal{O}|.
  \label{eq:cov_ratio}
\end{equation}

\subsubsection{Capacity Objective}\label{subsec:capacity}
Existing deployment optimization methods such as AutoBS~\cite{lee_autobs_2025} and AutoPlan~\cite{11587682} use a simplified SNR-based capacity metric derived from the RSS maps, expressed as
\begin{equation}
 C_{\mathrm{proxy}}(\mathcal B|\boldsymbol\psi)=\sum_{(i,j)\in\mathcal O}\log_2\!\left(1+\frac{\max_{m\in\mathcal B}10^{\psi_m(i,j)/10}}{N_0}\right),
 \label{eq:capacity_proxy}
\end{equation}
where $N_0$ is the thermal noise power in mW and RSS values are expressed in dBm. This metric omits inter-BS interference and user-dependent resource sharing, and can therefore overestimate capacity in dense deployments or heavily loaded cells. In this work, we instead account for both interference and spatial user demand through cell association and equal-share bandwidth allocation.

Beyond coverage, system capacity characterizes the cumulative achievable throughput the network can deliver. We quantify this objective through a per-pixel signal quality analysis associated with user spatial distribution. Under the max-RSS principle, each pixel $(i,j) \in \mathcal{O}$ is associated with the BS providing the strongest signal as
\begin{equation}
  \mathcal{C}(i,j,\mathcal{B}|\boldsymbol{\psi}) = \arg\max_{m \in \mathcal{B}}\ \psi_{m}(i,j).
  \label{eq:cell_assoc}
\end{equation}
Here, the cell association scheme $\mathcal{C}$ is a mapping that links pixel $(i,j)$ to its serving BS. It partitions outdoor areas $\mathcal{O}$ into $M$ disjoint serving regions, where the partition $\Omega_m^{\mathcal{B}} = \{(i,j) \in \mathcal{O} \mid \mathcal{C}(i,j,\mathcal{B}|\boldsymbol{\psi}) = m\}$ denotes the set of pixels served by BS $m$. For notational conciseness, we abbreviate the serving BS of pixel $(i,j)$ determined by $\mathcal{C}(i,j,\mathcal{B}|\boldsymbol{\psi})$ as $c_{ij}\in \mathcal{B}$ in subsequent equations when $\mathcal{B}$ and $\boldsymbol{\psi}$ are clear.

For pixel $(i,j) \in \mathcal{O}$ associated with its serving BS $c_{ij}$, the signal-to-interference-plus-noise ratio (SINR) is defined as
\begin{equation}
  \text{SINR}(i,j,\mathcal{B}|\boldsymbol{\psi}) = \frac{10^{\psi_{c_{ij}}(i,j)/10}}{\sum\nolimits_{{m \in \mathcal{B},\, m \neq c_{ij}}} 10^{\psi_{m}(i,j)/10} + N_0},
  \label{eq:sinr}
\end{equation}
where the numerator is the received signal power from the serving BS $c_{ij}$, and the denominator captures the aggregate interference combined with the thermal noise power $N_0$.

Assuming equal-share bandwidth allocation, the per-user bandwidth within the serving BS $m$ is 
\begin{equation}
  \tilde{B}_m^{\mathcal{B}} = B\big/\!\sum\nolimits_{(i,j) \in \Omega_m^{\mathcal{B}}} U_t(i,j).
  \label{eq:bandwidth}
\end{equation}

To account for practical modulation and coding schemes (MCS), spectral efficiency is modeled using Shannon's capacity formula and capped at a maximum value $\eta_{\max}$. This gives
\begin{equation}
  \eta(i,j,\mathcal{B}|\boldsymbol{\psi}) = \min\!\left\{\log_2\left(1 + \text{SINR}(i,j,\mathcal{B}|\boldsymbol{\psi})\right),\, \eta_{\max}\right\}.
  \label{eq:se_clip}
\end{equation}
From~\eqref{eq:bandwidth} and~\eqref{eq:se_clip}, the achievable data rate for a user at pixel $(i,j)$ is
\begin{equation}
  r(i,j,\mathcal{B}|\boldsymbol{\psi}) = \tilde{B}_{c_{ij}}^{\mathcal{B}} \cdot \eta(i,j,\mathcal{B}|\boldsymbol{\psi}).
  \label{eq:rate}
\end{equation}
Then the network capacity, defined as the cumulative throughput across all users, is expressed as
\begin{equation}
  R_{\text{tpt}}(\mathcal{B}|\boldsymbol{\psi}) = \sum\nolimits_{(i,j) \in \mathcal{O}} U_t(i,j) \cdot r(i,j,\mathcal{B}|\boldsymbol{\psi}).
  \label{eq:throughput}
\end{equation}

\subsection{Optimization Problem}\label{subsec:opt}

Since the user distribution $U_t(i,j)$ varies dynamically due to mobility, we formulate the capacity objective in terms of its expectation. For a given deployment $\mathcal{B}$, the SINR is deterministic, so only $U_t(i,j)$ introduces stochasticity. Expanding $r(i,j,\mathcal{B}|\boldsymbol{\psi})$ within~\eqref{eq:throughput} yields
\begin{align}
  R_{\text{tpt}}(\mathcal{B}|\boldsymbol{\psi}) = \sum_{(i,j) \in \mathcal{O}} \frac{U_t(i,j)}{\sum\nolimits_{(i',j') \in \Omega_{c_{ij}}^{\mathcal{B}}} U_t(i',j')} \cdot B \cdot \eta(i,j,\mathcal{B}|\boldsymbol{\psi}),
  \label{eq:thr_expanded}
\end{align}
which, by normalizing $U_t(i,j)$ with the total user count $N_\text{u}$, depends only on the user distribution $p_t(i,j) = U_t(i,j)/N_\text{u}$ rather than on the absolute user count. We define the expected user distribution $\rho(i,j) \triangleq \mathbb{E}_t[p_t(i,j)]$ and, by replacing $U_t(i,j)$ in the summand of~\eqref{eq:thr_expanded} with $\rho(i,j)$, define the contribution of each pixel to the expected throughput as
\begin{align}
 \bar{r}(i,j,\mathcal{B}|\boldsymbol{\psi},\boldsymbol{\rho}) = \frac{\rho(i,j)}{\sum_{(i',j') \in \Omega_{c_{ij}}^{\mathcal{B}}}\! \rho(i',j')} \! \cdot\! B\! \cdot \!\eta(i,j,\mathcal{B}|\boldsymbol{\psi}).
  \label{eq:r_expected}
\end{align}
Invoking the law of large numbers under the considered high-load regime, we approximate the expected throughput as
% \begin{align}
%   \mathbb{E}_t[R_{\text{tpt}}(\mathcal{B}|\boldsymbol{\psi})] \approx&\!\! \sum_{(i,j) \in \mathcal{O}} \frac{\rho(i,j)}{\sum_{(i',j') \in \Omega_{c_{ij}}^{\mathcal{B}}} \rho(i',j')}  \cdot B \cdot \eta(i,j,\mathcal{B}|\boldsymbol{\psi}) \notag \\
%   \triangleq&\ \bar{R}_{\text{tpt}}(\mathcal{B}|\boldsymbol{\psi},\boldsymbol{\rho}),
%   \label{eq:thr_expected}
% \end{align}
\begin{align}
  \mathbb{E}_t[R_{\text{tpt}}(\mathcal{B}|\boldsymbol{\psi})] \approx& \sum_{(i,j) \in \mathcal{O}} \bar{r}(i,j,\mathcal{B}|\boldsymbol{\psi},\boldsymbol{\rho})  \triangleq  \bar{R}_{\text{tpt}}(\mathcal{B}|\boldsymbol{\psi},\boldsymbol{\rho}).  \label{eq:thr_expected}
\end{align}
Here, the expected user distribution $\rho(i,j)$ can be estimated from real-world or synthetically generated user trajectory datasets.

Finally, the BS deployment problem is defined as selecting the optimal subset $\mathcal{B}^*$ that maximizes a weighted combination of coverage and expected system capacity. We define the objective as
\begin{equation}
  R(\mathcal{B}|\boldsymbol{\psi},\boldsymbol{\rho}) = \beta \cdot R_{\text{cov}}(\mathcal{B}|\boldsymbol{\psi}) + (1-\beta) \cdot \frac{\bar{R}_{\text{tpt}}(\mathcal{B}|\boldsymbol{\psi},\boldsymbol{\rho})}{R_{\text{norm}}},
  \label{eq:objective_R}
\end{equation}
and formulate the problem as
  \begin{align}\label{eq:opt}
    \mathcal{B}^* = \arg\max\nolimits_{\mathcal{B}} R(\mathcal{B}|\boldsymbol{\psi},\boldsymbol{\rho}),     \quad      \text{s.t.}        \  \mathcal{B} \subseteq \mathcal{Z}, \  |\mathcal{B}| = M,                                                                    
  \end{align}
where $\beta \in [0, 1]$ is a weighting factor that balances the coverage and capacity objectives, and $R_{\text{norm}}$ is a fixed normalization constant that scales the throughput to a comparable range. Larger $\beta$ favors coverage and smaller $\beta$ favors capacity.

Solving optimization problem~\eqref{eq:opt} for real-world BS deployment faces two critical challenges. First, calculation of the objective function requires site-specific radio maps $\boldsymbol{\psi}$ and the realistic user spatial density $\boldsymbol{\rho}$, both of which are difficult to obtain prior to deployment. Second, even with this information available, optimization problem~\eqref{eq:opt} is a combinatorial subset selection problem: choosing $M$ BSs from $K$ candidates yields $\binom{K}{M}$ possible solutions, where $K$ could exceed $10^4$ in kilometer-scale urban areas. Moreover, the nonconvex dependence of both coverage and capacity on BS locations, arising from complex propagation phenomena, cell association dynamics, and inter-cell interference patterns, precludes the application of standard gradient-based methods.

\section{Geographic Data-Informed Wireless Network Digital Twin}\label{sec:dt}

\begin{figure}[!t]
\centering
\includegraphics[width=0.95\linewidth]{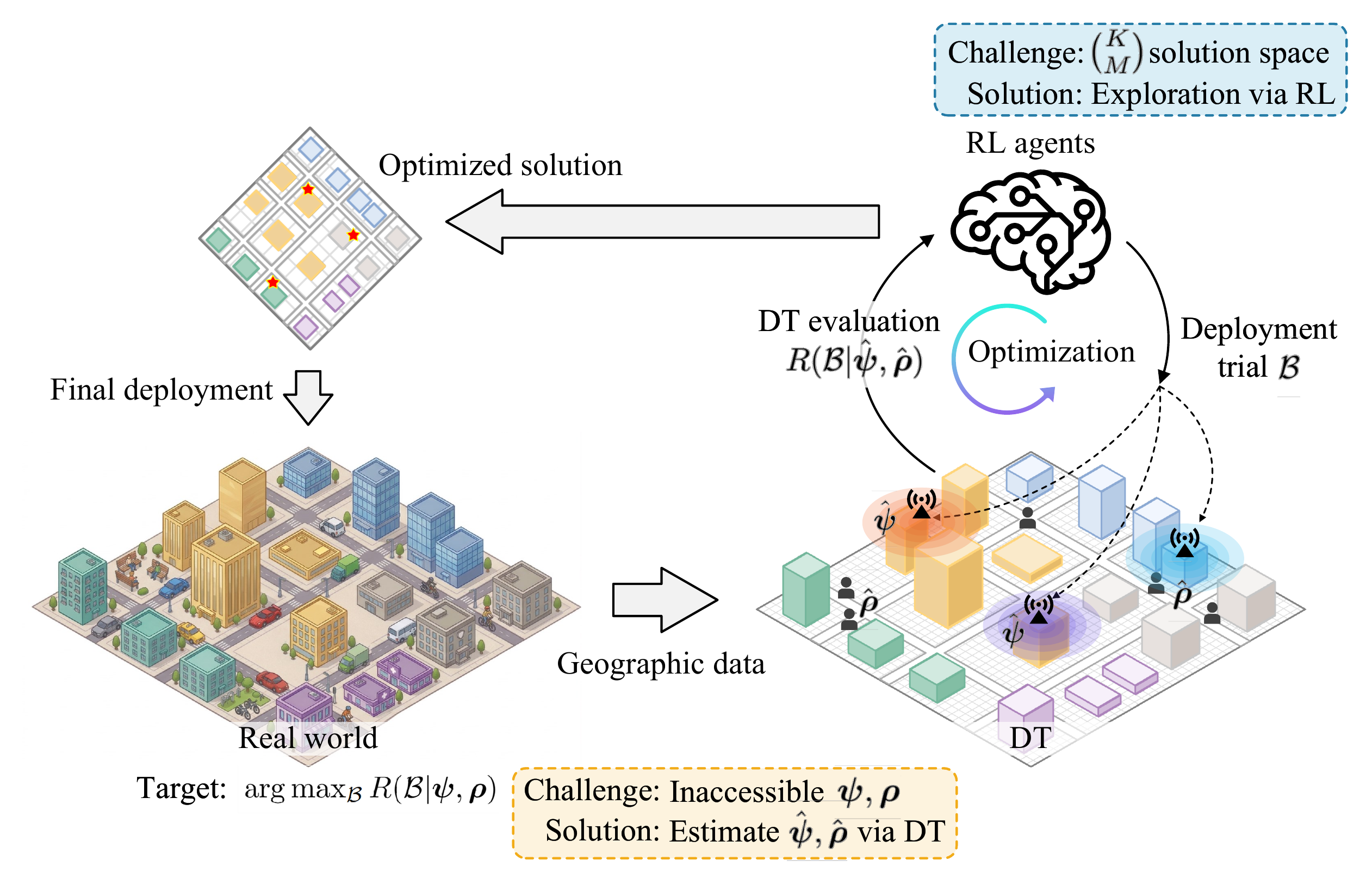}
\caption{Schematic of the proposed BS deployment framework.}
\label{fig:frame}
\end{figure}

To tackle the two challenges identified in Section~\ref{subsec:opt}, we propose an intelligent BS deployment framework, as illustrated in Fig.~\ref{fig:frame}. For the first challenge, we construct a geographic data-informed wireless network DT that estimates $\hat{\boldsymbol{\psi}}$ and $\hat{\boldsymbol{\rho}}$ directly from open geographic data. The DT integrates a sample-free radio map prediction model that captures complex propagation across diverse building heights and heterogeneous materials at the kilometer scale, and a generative user mobility model that synthesizes road topology-constrained trajectories and reflects realistic spatial traffic demand. This yields the DT-based objective $R(\mathcal{B}|\hat{\boldsymbol{\psi}},\hat{\boldsymbol{\rho}})$ that enables performance evaluation at the millisecond level for any deployment candidate $\mathcal{B}$. For the second challenge, we develop a DRL-based optimization module that leverages the DT environment to efficiently identify high-quality deployments within a vast combinatorial solution space. The remainder of this section details the construction of the geographic data-informed DT.

\subsection{Sample-Free Radio Map Prediction Model}\label{subsec:radiomap}

\begin{figure}[!t]
  \centering
  \includegraphics[width=0.95\linewidth]{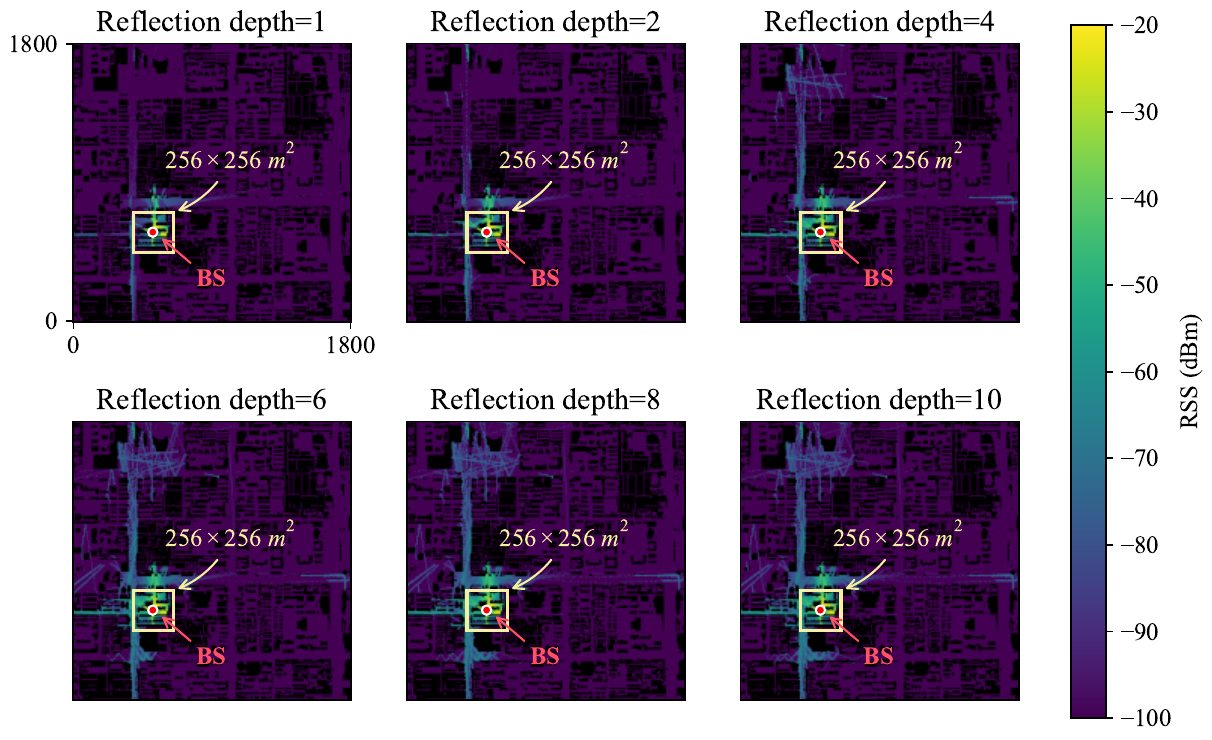}
  \caption{Comparison of RT radio maps under different maximum reflection depths (1-10) for a representative area.}
  \label{fig:reflection_depth}
\end{figure}

Evaluating the deployment objective in~\eqref{eq:opt} requires site-specific radio maps for every candidate BS location, yet acquiring $\boldsymbol{\psi}$ through field measurements or on-demand RT incurs prohibitive cost. Sample-free radio map prediction, which infers propagation characteristics directly from environmental descriptions, offers a scalable alternative. However, existing methods operate on compact regions, typically $256\times256$~$\text{m}^2$, where building height variations are limited and signals attenuate within one to two reflections~\cite{jaensch_radio_2025}. These conditions permit simplifying 3D geometry into 2D binary maps and neglecting material-specific EM properties~\cite{lee2023pmnet,lee_scalable_2024,11174709}. At the kilometer scale relevant to macro BS deployment, building heights vary significantly, creating over-rooftop diffraction and height-dependent reflection paths that 2D representations cannot capture. Meanwhile, multiple reflection orders provide non-negligible contributions to the RSS from macro BSs, as illustrated in Fig.~\ref{fig:reflection_depth}, amplifying the influence of heterogeneous materials on propagation loss. To address these limitations, we first construct a new kilometer-scale dataset that incorporates 3D urban morphology and material-specific EM properties tailored for macro BS deployment, and then develop a prediction model that estimates site-specific radio maps directly from the geographic data.

\subsubsection{Radio Map Dataset Construction}

We extract building footprints from OpenStreetMap for multiple $1800 \times 1800$~$\text{m}^2$ urban areas, discretized into a $256\times256$ pixel grid. While the footprint geometry is readily available, height annotations are often incomplete. For buildings with available height tags, we directly adopt the annotated values. For the remainder, we develop a density-aware stochastic generation method. Specifically, we compute the local building density, defined as the ratio of building area to the total neighborhood area, and classify buildings into four morphology categories, including financial district, high-rise residential area, mixed-use area, and old town area. The height $h$ of a building in category $c$ is then sampled from $ h \sim \text{LogNormal}(\mu_c, \sigma_c^2)$, where $\mu_c$ and $\sigma_c$ are respectively the category-specific log-scale mean and standard deviation, and is clipped between 6~m and 300~m to ensure physically plausible urban building heights.

To incorporate diversified EM properties, we assign heterogeneous material types to buildings based on the ITU-R P.2040 recommendation~\cite{itu_p2040}. Four candidate building materials and a ground material type are considered, each characterized by its relative permittivity $\varepsilon_r$ and conductivity $\sigma$, as summarized in Table~\ref{table:materials}.
\begin{table}[!t]
  \centering
  \caption{Building Material EM Properties at 3.5~GHz~\cite{itu_p2040}}
  \label{table:materials}
  \renewcommand{\arraystretch}{0.65}
  \begin{tabular}{lcc}
    \toprule
    \textbf{Material Type}    & \textbf{Relative Permittivity $\varepsilon_r$} & \textbf{Conductivity $\sigma$ (S/m)} \\
    \midrule
    Glass                     & 6.310                                          & 0.0036                               \\
    Concrete                  & 5.240                                          & 0.0462                               \\
    Brick                     & 3.910                                          & 0.0238                               \\
    Marble                    & 7.074                                          & 0.0055                               \\
    Very dry ground           & 3.000                                          & 0.00015                              \\
    \bottomrule
  \end{tabular}
\end{table}
To simulate realistic urban conditions where buildings within the same commercial complex or residential compound tend to share similar construction materials, we adopt a building grouping strategy. Adjacent buildings whose mutual distance falls below a predefined threshold are clustered into a group that shares the same material type and height generation process. This grouping mechanism captures the spatial coherence of construction in real-world urban environments. Based on the above geographic data processing pipeline, Sionna~\cite{sionna} is employed to generate site-specific radio maps through RT.

% Reproducibility: confirm the dataset-generation version before reporting its
% density-to-lognormal parameters, density window, grouping threshold, and material
% assignment probabilities. Current generator defaults alone do not establish the
% settings used for the published dataset.
\subsubsection{Hybrid Input Representation}
\begin{figure*}[!t]
  \centering
  \includegraphics[width=0.95\linewidth]{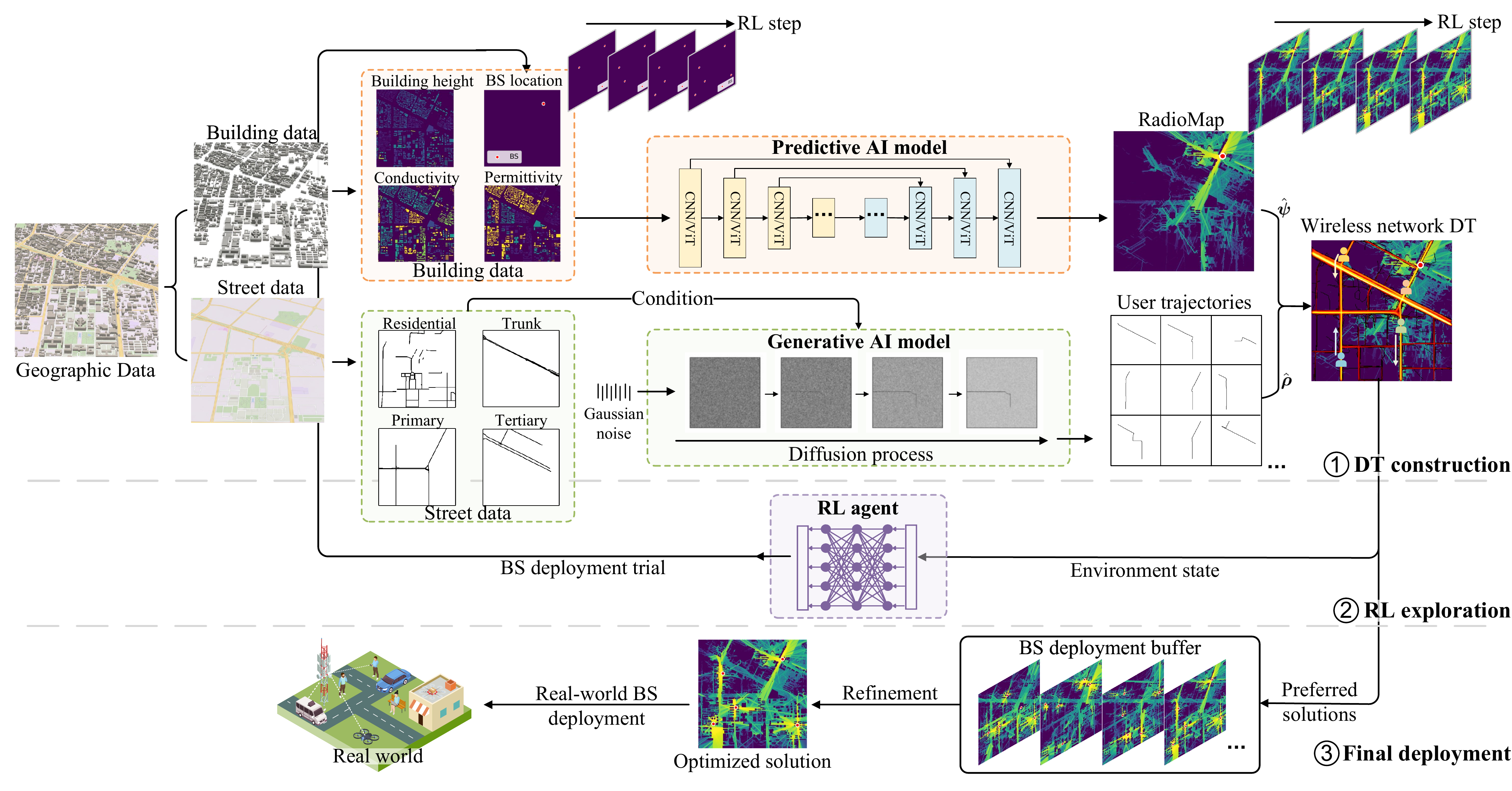}
  \caption{Intelligent BS deployment framework based on geographic data-informed DT.}
  \label{fig:dt_framework}
\end{figure*}

Moving beyond simplified 2D building maps, the proposed model employs a hybrid input representation to capture diffraction and high-order reflections in urban scenarios. As shown in the upper part of Fig.~\ref{fig:dt_framework}, the building information of a specific area is encoded into a four-channel image, denoted by $\mathbf{X} \in \mathbb{R}^{4 \times N_{\text{p}} \times N_{\text{p}}}$, where each channel captures a distinct physical attribute:
\begin{itemize}
  \item \textbf{Channel 1: 3D building height map.} Each pixel records the building height at that location with zeros for outdoor pixels, capturing vertical propagation effects such as diffraction and reflections.
  \item \textbf{Channel 2: BS map.} The transmitter location and antenna height are encoded, enabling the model to handle variable BS heights.
  \item \textbf{Channels 3-4: Conductivity and permittivity maps.} Each building pixel is assigned the conductivity $\sigma$ and relative permittivity $\varepsilon_r$ in two separate channels, encoding the material-dependent EM properties.
\end{itemize}
The output is a single-channel RSS map $\hat{\mathbf{Y}} \in \mathbb{R}^{1 \times N_{\text{p}} \times N_{\text{p}}}$, where each pixel $(i,j) \in \mathcal{O}$ encodes the predicted RSS $\hat{\psi}_m(i,j)$ in dBm. This hybrid representation enables the model to jointly reason about 3D geometric obstruction, transmitter configuration, and EM interactions. Notably, it is also extensible to sample-based prediction by, e.g., appending two additional channels, i.e., a binary sample mask and corresponding measured signal strengths, while remaining fully functional in the sample-free regime.

\subsubsection{Network Architecture and Data Augmentation}
The proposed hybrid input representation is architecture-agnostic and can be integrated with existing radio map prediction backbones. To validate this generality, we evaluate three representative models, specifically U-Net~\cite{ronneberger2015u}, PMNet~\cite{lee2023pmnet}, and vision transformer (ViT)-based RMTransformer~\cite{11174709}. To enhance generalization, input-target pairs $(\mathbf{X}, \mathbf{Y})$, where $\mathbf{Y}$ denotes the ground-truth RSS map, are randomly subjected to one of eight geometric transformations during training, comprising four $90^\circ$ rotations and their horizontal mirrors, applied identically to all channels. This augmentation multiplies the effective training set size eightfold while preserving physical consistency between environment representations and the RSS maps.

Through the above design, the trained model can predict the radio map $\hat{\boldsymbol{\psi}}$ for any candidate BS location at the millisecond level, given only the openly available geographic data.

\subsection{Generative User Mobility Model}\label{subsec:mobility}

Beyond radio maps, the deployment objective in~\eqref{eq:thr_expected} requires the expected user spatial distribution $\boldsymbol{\rho}$ to evaluate the capacity objective. Existing trajectory generation methods~\cite{9338370,DiffTraj} can produce realistic mobility traces, yet they rely on substantial real trajectory datasets collected in specific target areas. Such data are typically unavailable prior to network deployment, and their collection raises significant privacy concerns, making them unsuitable for estimating $\hat{\boldsymbol{\rho}}$. To eliminate this dependency on real trajectory data from the target area during inference, we introduce Map2Traj, a zero-shot generative user mobility model built upon the denoising diffusion probabilistic model (DDPM)~\cite{NEURIPS2020_4c5bcfec}. By conditioning solely on hierarchical street map images extracted from OpenStreetMap~\cite{haklay2008openstreetmap}, Map2Traj synthesizes road topology-constrained trajectories for arbitrary target areas, enabling the estimation of $\hat{\boldsymbol{\rho}}$ from openly available geographic data alone.

\subsubsection{Diffusion-Based Trajectory Generation}
Map2Traj represents street maps as multi-channel images and trajectories as single-channel images. The street map $\boldsymbol{\lambda}$ is constructed by categorizing road segments into multiple hierarchical levels based on their OpenStreetMap tags, such as \textit{Trunk}, \textit{Primary}, and \textit{Residential}, and encoding each level as a separate binary channel, thereby preserving the distinct characteristics of different road types. Individual trajectories are represented as single-channel binary images $\boldsymbol{l}$.

The generative process consists of a forward diffusion and a reverse denoising procedure. The forward diffusion process adheres to the standard DDPM paradigm, which iteratively adds Gaussian noise to a real trajectory image $\boldsymbol{l}_0$ over $T$ time steps. Specifically, at each step $t$, the noisy trajectory $\boldsymbol{l}_t$ is sampled from
\begin{equation}
  q\left(\boldsymbol{l}_t | \boldsymbol{l}_{t-1}\right) = \mathcal{N}\left(\boldsymbol{l}_t;\, \sqrt{\alpha_t}\, \boldsymbol{l}_{t-1},\, (1-\alpha_t) \mathbf{I}\right),
  \label{eq:forward_step}
\end{equation}
where $\alpha_t$ for $t=1,2,\dots,T$ are hyperparameters of the noising schedule, $\mathcal{N}(\boldsymbol{x};\boldsymbol{\mu},\boldsymbol{\nu})$ denotes the normal distribution with mean $\boldsymbol{\mu}$ and covariance $\boldsymbol{\nu}$, and $\mathbf{I}$ is the identity matrix. This process progressively noises the real trajectory $\boldsymbol{l}_0$ until $\boldsymbol{l}_T$ becomes indistinguishable from standard Gaussian noise at the $T$-th step. Leveraging the additive properties of Gaussian distributions, the forward process can also be marginalized to directly obtain the noisy trajectory at an arbitrary step $t$ as:
\begin{equation}
  q\left(\boldsymbol{l}_t | \boldsymbol{l}_0\right) = \mathcal{N}\left(\boldsymbol{l}_t;\, \sqrt{\gamma_t}\, \boldsymbol{l}_0,\, (1-\gamma_t) \mathbf{I}\right),
  \label{eq:forward}
\end{equation}
where $\gamma_t = \prod_{i=1}^{t} \alpha_i$. This closed-form expression enables efficient sampling during training. The reverse process then recovers trajectories from Gaussian noise by learning a conditional denoising distribution:
\begin{equation}
  p_{\boldsymbol{\theta}}(\boldsymbol{l}_{0:T} | \boldsymbol{\lambda}) = p(\boldsymbol{l}_T) \prod\nolimits_{t=1}^{T} p_{\boldsymbol{\theta}}(\boldsymbol{l}_{t-1} | \boldsymbol{l}_t, \boldsymbol{\lambda}),
  \label{eq:reverse}
\end{equation}
where $p(\boldsymbol{l}_T) = \mathcal{N}(\boldsymbol{l}_T; \mathbf{0}, \mathbf{I})$, $p_{\boldsymbol{\theta}}$ is a learnable distribution parameterized by $\boldsymbol{\theta}$. A neural network $f_{\boldsymbol{\theta}}(\boldsymbol{\lambda}, \boldsymbol{l}_t, t)$ is trained to predict the noise component from the concatenation of $\boldsymbol{\lambda}$ and the noisy trajectory $\boldsymbol{l}_t$, with the denoising step given by
\begin{equation}
  \boldsymbol{l}_{t-1} \leftarrow \frac{1}{\sqrt{\alpha_t}} \left( \boldsymbol{l}_t - \frac{1-\alpha_t}{\sqrt{1-\gamma_t}} f_{\boldsymbol{\theta}}(\boldsymbol{\lambda}, \boldsymbol{l}_t, t) \right) + \sqrt{1-\alpha_t}\, \boldsymbol{\epsilon},
  \label{eq:denoise_step}
\end{equation}
where $\boldsymbol{\epsilon} \sim \mathcal{N}(\mathbf{0}, \mathbf{I})$. The key advantage of Map2Traj is that the condition $\boldsymbol{\lambda}$ provides continuous spatial guidance throughout the entire denoising process, ensuring that the model generates trajectories that follow the hierarchical street topology in target areas, including areas unseen during training.

\subsubsection{Estimation of User Spatial Distribution}
Given the street map $\boldsymbol{\lambda}$ of the target deployment area, Map2Traj generates a set of $N_{\text{traj}}$ synthetic trajectories, denoted by $\{\hat{\boldsymbol{l}}^{(1)}, \hat{\boldsymbol{l}}^{(2)}, \ldots, \hat{\boldsymbol{l}}^{(N_{\text{traj}})}\}$. The estimated distribution $\hat{\boldsymbol{\rho}}$ is calculated by aggregating and normalizing these trajectories:
\begin{equation}
  \hat{\rho}(i,j) = \frac{\sum_{n=1}^{N_{\text{traj}}} \hat{\boldsymbol{l}}^{(n)}(i,j)}{\sum_{(i',j') \in \mathcal{O}} \sum_{n=1}^{N_{\text{traj}}} \hat{\boldsymbol{l}}^{(n)}(i',j')},
  \label{eq:rho_est}
\end{equation}
where $\hat{\boldsymbol{l}}^{(n)}(i,j)$ indicates the presence of the $n$-th trajectory at pixel $(i,j)$. This estimation process requires only the street map from OpenStreetMap and uses neither real target-area user trajectories nor privacy-sensitive information during inference.

By combining the radio map prediction model and the generative user mobility model, we construct a complete geographic data-informed wireless network DT, as illustrated in the upper part of Fig.~\ref{fig:dt_framework}. Notably, both $\hat{\boldsymbol{\psi}}$ and $\hat{\boldsymbol{\rho}}$ originate from the same geographic data source, where building data governs radio propagation and the associated street layout constrains user mobility, thereby establishing a coherent coupling between channel conditions and user spatial distribution within a unified DT. With $\hat{\boldsymbol{\psi}}$ and $\hat{\boldsymbol{\rho}}$ readily available from AI models, the coverage ratio $R_{\text{cov}}(\mathcal{B}|\hat{\boldsymbol{\psi}})$ and expected throughput $\bar{R}_{\text{tpt}}(\mathcal{B}|\hat{\boldsymbol{\psi}},\hat{\boldsymbol{\rho}})$ in~\eqref{eq:opt} can be evaluated at the millisecond level for any deployment candidate $\mathcal{B}$, which allows efficient exploration of the combinatorial solution space through the DRL-based optimization framework presented in the next section.

\section{DRL-Based Intelligent BS Deployment via Wireless Network Digital Twin}\label{sec:drl}

Building upon the wireless network DT in Section~\ref{sec:dt}, this section develops a DRL-based optimization framework to solve the combinatorial BS deployment problem in~\eqref{eq:opt}. We formulate the deployment task as a sequential decision-making problem as an MDP and present a spatially structured PPO agent enhanced with LS and a deployment buffer, as illustrated in the lower part of Fig.~\ref{fig:dt_framework}. 
%maintain the diversity of explored deployment configurations.

\subsection{MDP Formulation}\label{subsec:mdp}

We model the BS deployment process as a multi-step MDP to select $M$ BS locations from the candidate set $\mathcal{Z}$, with the DT evaluating network performance after each placement. The MDP is defined by the tuple $(\mathcal{S}, \mathcal{A}, \mathcal{R}, \mathcal{P}, \gamma)$, denoting the state space, action space, reward function, state transitions, and discount factor, respectively.

\subsubsection{State Space}
The state $\mathbf{s}_n$ at step $n$ is composed of five spatial feature maps, each of size $N_{\text{p}} \times N_{\text{p}}$:
\begin{equation}
  \mathbf{s}_n = \left(\mathbf{S}_n^{\text{c}},\; \mathbf{S}_n^{\text{d}},\; \mathbf{S}_n^{\text{r}},\;\mathbf{S}_n^{\text{cov}},\;\mathbf{S}_n^{\text{rate}}\right) \in \mathbb{R}^{5 \times N_{\text{p}} \times N_{\text{p}}}.
  \label{eq:state}
\end{equation}

\textit{Candidate BS map} $\mathbf{S}_n^{\text{c}}$: A binary map of available candidate locations, initialized with all $K$ positions in $\mathcal{Z}$. To enforce a minimum inter-BS separation, deploying a BS at $z_m$ removes all candidates within an inhibition radius $r_{\text{inh}}$:
\begin{equation}
  \mathbf{S}_{n+1}^{\text{c}}(i,j) = \mathbf{S}_n^{\text{c}}(i,j) \cdot \mathbb{I}\left(\|(i,j) - (x_m,y_m)\|_2 > r_{\text{inh}}\right),
  \label{eq:inhibition}
\end{equation}
where $(x_m,y_m)$ are horizontal coordinates of the newly deployed BS $z_m$. Coordinates and $r_{\text{inh}}$ are expressed in grid units.

\textit{Deployed BS map} $\mathbf{S}_n^{\text{d}}$: A binary map recording the positions of all deployed BSs, defined as $\mathbf{S}_n^{\text{d}}(i,j) = \mathbb{I}\left((i,j) \in \{(x_m,y_m) \mid z_m\in\mathcal{B}_{n-1}\}\right)$, where $\mathcal{B}_{n-1}$ denotes the BSs deployed after step $n-1$. This channel enables the policy to reason about spatial complementarity with existing BSs.

\textit{Existing radio map} $\mathbf{S}_n^{\text{r}}$: The composite radio coverage from all previously deployed BSs, computed as the pixel-wise maximum RSS:
\begin{equation}
  \mathbf{S}_n^{\text{r}}(i,j) = \max_{m \in \mathcal{B}_{n-1}} \hat{\psi}_{m}(i,j).
  \label{eq:composite_radio}
\end{equation}

\textit{Coverage map} $\mathbf S_n^{\text{cov}}$: A binary map marking outdoor pixels whose maximum RSS reaches $\psi_{\text{thr}}$, explicitly indicating coverage gaps.

\textit{Expected-rate map} $\mathbf S_n^{\text{rate}}$: The pixel-wise expected throughput contribution defined in~\eqref{eq:r_expected}, evaluated with the current deployed set $\mathcal B_{n-1}$ and DT estimates $\hat{\boldsymbol\psi},\hat{\boldsymbol\rho}$.
% Pending implementation: the current fifth channel is normalized spectral
% efficiency, not eq:state_rate. Do not relabel existing results as this version.

\subsubsection{Action Space}
The action $a_n$ at step $n$ corresponds to selecting a single pixel $(x, y)$ from the $N_{\text{p}} \times N_{\text{p}}$ grid as the next BS location:
\begin{equation}
  a_n = (x_n, y_n) \in \{1, \ldots, N_{\text{p}}\}^2.
  \label{eq:action}
\end{equation}
To ensure that only valid candidate positions are selected, we apply the candidate BS map $\mathbf{S}_n^{\text{c}}$ to filter the policy output. The selected pixel $a_n=(x_n,y_n)$ corresponds to a candidate $z(a_n)=(x_n,y_n,h_n)\in\mathcal Z$, where $h_n$ is its antenna height.

% Specifically, the policy network outputs per-pixel logits $\boldsymbol{\ell}(\mathbf{s}_n) \in \mathbb{R}^{N_{\text{p}} \times N_{\text{p}}}$, and the masked policy distribution is obtained by setting the logits of invalid pixels to $-\infty$ before applying the softmax operation:
% \begin{equation}
%   \pi_{\boldsymbol{\theta}}(a_n | \mathbf{s}_n) = \frac{\exp\left(\boldsymbol{\ell}_{a_n}(\mathbf{s}_n)\right)}{\sum_{a' \in \mathcal{A}_n^{\text{valid}}} \exp\left(\boldsymbol{\ell}_{a'}(\mathbf{s}_n)\right)},
%   \label{eq:masked_policy}
% \end{equation}
% where $\mathcal{A}_n^{\text{valid}} = \{(x,y) \mid \mathbf{S}_n^{\text{c}}(x,y) = 1\}$ denotes the set of valid actions at step $n$. This formulation ensures that every sampled action corresponds to a feasible candidate BS location while maintaining a fully differentiable policy.

\subsubsection{Reward Function}
We define the step-wise reward as the incremental objective gain from placing a new BS. For step $n \ge 2$, the reward is 
\begin{equation}
  R_n = R(\mathcal{B}_n|\hat{\boldsymbol{\psi}},\hat{\boldsymbol{\rho}}) - R(\mathcal{B}_{n-1}|\hat{\boldsymbol{\psi}},\hat{\boldsymbol{\rho}}),
  \label{eq:reward}
\end{equation}
 and for the first step we use $R_1 = R(\mathcal{B}_1|\hat{\boldsymbol{\psi}},\hat{\boldsymbol{\rho}})$. Here, $\mathcal{B}_n$ denotes the set of BSs placed after step $n$, $\hat{\boldsymbol{\psi}}$ and $\hat{\boldsymbol{\rho}}$ are the DT-predicted radio maps and user spatial distribution, respectively, and the objective $R(\cdot|\hat{\boldsymbol{\psi}},\hat{\boldsymbol{\rho}})$ follows the definition in \eqref{eq:objective_R}. The episode terminates immediately after the $M$-th BS is deployed.

\subsubsection{State Transition}
\begin{figure*}[!t]
 \centering
 \includegraphics[width=0.72\linewidth]{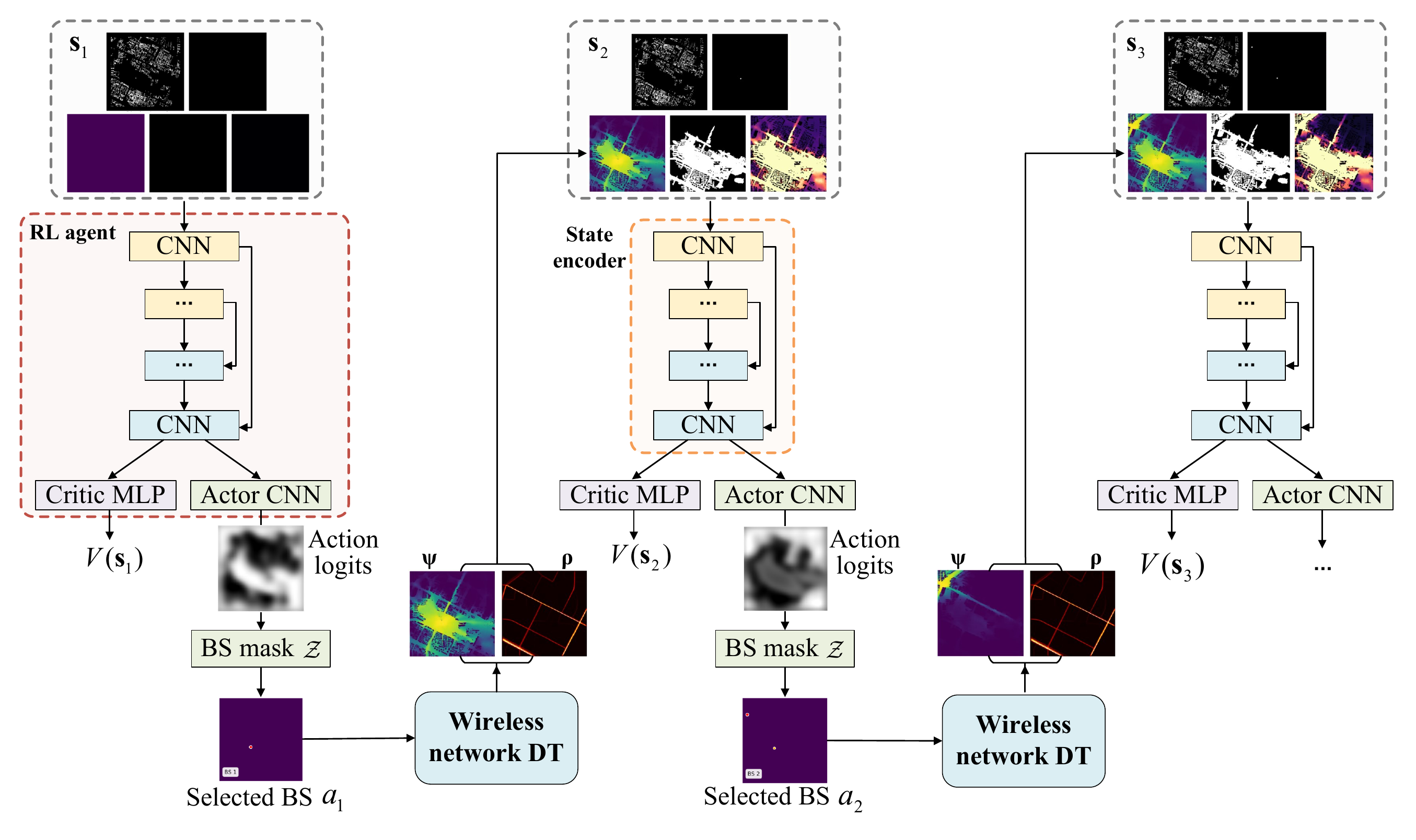}
 \caption{Spatial PPO structure and interaction with the wireless network DT.}
 \label{fig:drl_structure}
\end{figure*}

Given the current state $\mathbf{s}_n$ and action $a_n = (x_n, y_n)$, the transition to $\mathbf{s}_{n+1}$ is deterministic: the deployed BS map is updated by marking pixel $(x_n, y_n)$, the candidate map is updated via the inhibition rule in~\eqref{eq:inhibition}, and the existing radio map is recomputed using the wireless network DT to predict the signal strength for the newly placed BS and merge it with the existing composite map via~\eqref{eq:composite_radio}. Coverage and expected-rate maps are then recomputed for the updated deployment, as shown in Fig.~\ref{fig:drl_structure}.

This state representation also supports incremental deployment in an existing network by treating the installed BSs as an already completed portion of the MDP. Given $M_0<M$ existing BSs with deployment set $\mathcal B_{M_0}$, the DT constructs $\mathbf s_{M_0+1}$ from their deployed BS, composite RSS, coverage, and rate maps, and updates the candidate map to exclude occupied sites and their adjacent regions. The agent then continues from step $n=M_0+1$ to $M$, selecting the remaining $M-M_0$ BS locations with the existing BSs held fixed.

\subsection{Spatial PPO Model}\label{subsec:ppo}
PPO implementations typically employ multilayer perceptrons (MLPs) to extract state features, offering a simple and computationally efficient architecture for compact state representations. AutoBS~\cite{lee_autobs_2025} follows this design and evaluates deployment over $512\times512$~$\text{m}^2$ areas. For the kilometer-scale urban environments considered in this study, however, flattened map inputs would obscure spatial relationships and geometric dependencies among deployment locations, limiting effective capture of complex geographic structure. Dense connections also incur substantial parameter overhead for high-dimensional spatial states. Inspired by the radio map prediction models in Section~\ref{subsec:radiomap}, we adopt a U-Net model~\cite{ronneberger2015u} as a state encoder for spatial state feature extraction. Its structure aggregates spatial context at multiple scales, while skip connections preserve fine-grained location information for pixel-wise BS selection.

Fig.~\ref{fig:drl_structure} illustrates the shared state encoder and its two output branches at successive deployment steps. At step $n$, the five-channel state $\mathbf s_n$ passes through the state encoder once, and both actor and critic reuse the extracted spatial features, avoiding duplicated feature extraction during agent inference. Two lightweight heads then map the shared features to the outputs required by PPO: the critic uses global average pooling followed by an MLP head to produce the scalar state value $V_\theta(\mathbf s_n)$, whereas the actor uses a convolutional head to produce spatial action logits over the $256\times256$ action grid, followed by BS location masking via~\eqref{eq:inhibition} to obtain the policy $\pi_\theta(a_n|\mathbf s_n)$. The resulting spatial PPO architecture combines location-preserving action generation with shared policy and value feature extraction for efficient deployment search in the DT. The masked action selects the next BS, after which the DT updates the deployment state and returns the incremental reward, forming the sequential interaction shown in Fig.~\ref{fig:drl_structure}.

The agent is trained following the standard PPO procedure~\cite{schulman2017proximal}, using the clipped surrogate objective
\begin{equation}
 L^{\mathrm{clip}}(\theta)=\mathbb E_n\!\left[\min\!\left(\xi_n(\theta)\hat A_n,
 \operatorname{clip}(\xi_n(\theta),1-\epsilon,1+\epsilon)\hat A_n\right)\right],
 \label{eq:ppo_clip}
\end{equation}
where $\xi_n(\theta)=\frac{\pi_\theta(a_n|\mathbf s_n)}{\pi_{\theta_{\mathrm{old}}}(a_n|\mathbf s_n)}$
is the policy probability ratio with respect to the rollout policy with parameters $\theta_{\mathrm{old}}$, $\hat A_n=\sum_{k=n}^{M}\gamma^{k-n}R_k-V_\theta(\mathbf s_n)$ is the estimated advantage, $\epsilon$ is the clipping threshold, and $\mathbb E_n$ averages over rollout transitions. 

\subsection{Local Search Refinement}\label{subsec:local_search}

While the DRL agent learns to produce high-quality deployments in the DT, stochastic exploration may not traverse all locations in $\mathcal{Z}$. We therefore introduce an LS process that refines the DRL-generated deployment in a post-training phase.

Given a deployment $\mathcal{B} = \{z_1, z_2, \ldots, z_M\}$ retained during training, the LS process alternately optimizes each BS location while keeping the remaining $M-1$ BSs fixed. Specifically, for each BS $z_m \in \mathcal{B}$, a local candidate grid $\Omega_m$ is constructed, containing all candidate positions within a refinement radius $r_{\text{ref}}$ centered at $z_m$:
\begin{equation}
  \Omega_m = \{z \in \mathcal{Z} \mid \|z - z_m\|_2 \leq r_{\text{ref}}\}.
  \label{eq:local_grid}
\end{equation}
For each alternative position $\omega \in \Omega_m$, the DT rapidly evaluates the objective in~\eqref{eq:objective_R} for the modified deployment, defined by $\mathcal{B}_{z_m\leftarrow\omega} \triangleq \mathcal{B}\! \setminus\! \{z_m\}\! \cup\! \{\omega\}$. The BS $z_m$ is then relocated to the position yielding the highest objective:
\begin{equation}
  z_m^* = \arg\max\nolimits_{\omega \in \Omega_m}\; R(\mathcal{B}_{z_m\leftarrow\omega}|\hat{\boldsymbol{\psi}},\hat{\boldsymbol{\rho}}).
  \label{eq:local_opt}
\end{equation}
This process is repeated until no improvement is observed in a full round. The millisecond-level evaluation enabled by the DT makes repeated LS evaluations computationally tractable even for large candidate sets.

The LS process serves as a complementary exploitation mechanism to the DRL's global exploration. While the DRL agent efficiently navigates the exponentially large combinatorial space to identify promising deployment regions, the LS performs further fine-grained coordinate-level optimization within these regions. 

\subsection{Wasserstein Distance-Based Deployment Buffer}\label{buffer-section}

The quality of LS refinement depends on the starting deployment, since each search explores a local neighborhood of the initial BS configuration. Applying LS only to the best deployment encountered during DRL training may therefore overlook other promising regions of the combinatorial solution space. Retaining multiple high-reward deployments provides alternative starting points, but a buffer based solely on reward ranking may contain spatially similar configurations as the policy converges. These configurations may lead to the same local optimum, limiting the benefit of multiple LS runs.

To retain both deployment quality and spatial diversity, we introduce a Wasserstein distance-based deployment buffer $\mathcal Q$. The buffer stores at most $N_{\text{buf}}$ tuples $(\mathcal B,R(\mathcal B))$, where $\mathcal B=\{z_1,\ldots,z_M\}$ and $R(\mathcal B)$ abbreviates the DT objective $R(\mathcal B|\hat{\boldsymbol\psi},\hat{\boldsymbol\rho})$ in~\eqref{eq:objective_R}. Starting from an empty buffer, completed deployments are selectively retained during training according to their objective values and spatial dissimilarity, providing multiple starting points for subsequent LS refinement.

\subsubsection{Wasserstein Distance-Based Diversity Metric}
To quantify the structural dissimilarity between deployments, we treat each as an unordered set of 3D BS coordinates. The discrete Wasserstein-1 distance is
\begin{equation}
 W(\mathcal B_1,\mathcal B_2)=\frac1M\min_{\sigma\in\Pi_M}\sum_{m=1}^M
 \|z_m^{(1)}-z_{\sigma(m)}^{(2)}\|_2,
 \label{eq:wasserstein}
\end{equation}
where $\Pi_M$ is the set of permutations of $\{1,\ldots,M\}$, and $z_m^{(1)}$ and $z_{\sigma(m)}^{(2)}$ denote the coordinates of the $m$-th BS in $\mathcal B_1$ and its matched counterpart in $\mathcal B_2$, respectively. By minimizing the average distance over all one-to-one matchings, this metric compares deployment geometry independently of BS ordering. A smaller value indicates greater spatial similarity. The optimal assignment is computed by the Hungarian algorithm~\cite{munkres1957algorithms} in $O(M^3)$ time, making the comparison tractable for the considered BS deployment scales.

\subsubsection{Diversity-Preserving Update Rule}\label{subsubsec:buffer_update}
At the end of each training episode, the completed deployment and its DT objective, $(\mathcal B^{\text{new}},R(\mathcal B^{\text{new}}))$, are considered for insertion into $\mathcal{Q}$. If $\mathcal Q$ is empty, this tuple is directly appended. Otherwise, the nearest buffered deployment is identified by $q^*=\arg\min_q W(\mathcal B^{\text{new}},\mathcal B^{(q)})$, where $\mathcal B^{(q)}$ denotes the $q$-th stored deployment. A predefined similarity threshold $\delta$ determines which of the following rules is applied:
\begin{itemize}
  \item \textit{Similar deployment update:} If $W(\mathcal B^{\text{new}},\mathcal B^{(q^*)})<\delta$, the new deployment replaces $\mathcal B^{(q^*)}$ only when $R(\mathcal B^{\text{new}})>R(\mathcal B^{(q^*)})$. This retains the better solution among the two spatial neighbors.
  \item \textit{Diverse deployment insertion:} If $W(\mathcal B^{\text{new}},\mathcal B^{(q^*)})\geq\delta$, the new deployment is appended when $|\mathcal Q|<N_{\text{buf}}$. If the buffer is full, it replaces the lowest-reward entry only when its objective exceeds that entry's reward; otherwise, it is discarded.
\end{itemize}
Each replacement also updates the stored objective value, while rejected candidates leave the buffer unchanged. The threshold $\delta$ controls the spatial scale at which deployments are treated as similar: increasing it subjects more candidates to the local replacement rule, whereas decreasing it allows finer spatial distinctions during insertion.

Together, these components constitute a three-stage DT-based BS deployment framework that integrates global exploration, diversity-preserving selection, and multi-start local refinement. The DRL agent explores the combinatorial deployment space, the Wasserstein distance-based buffer selects promising and spatially distinct deployments, and LS exploits their local neighborhoods to obtain the final deployment with the highest refined DT objective. By coupling reward-driven exploration with deployment-level diversity selection, the framework directs local refinement toward multiple promising regions of the solution space, reducing its reliance on a single DRL-generated starting point. Algorithm~\ref{alg:deployment} summarizes the complete procedure.

\begin{algorithm}[!t]
\SetAlgoLined
\caption{DT-based intelligent BS deployment}\label{alg:deployment}
\KwIn{DT, $\mathcal Z$, $M$, $r_{\mathrm{inh}}$, $r_{\mathrm{ref}}$, $N_{\mathrm{buf}}$, $\delta$, and DRL training budgets.}
\KwOut{Deployment $\mathcal B^*$.}
Initialize agent $(\pi_\theta,V_\theta)$; deployment buffer $\mathcal Q\leftarrow\emptyset$\;
\tcc{1. PPO exploration within the DT}
\While{training episode budget remains}{
 $\theta_{\mathrm{old}}\leftarrow\theta$; initialize rollout storage $\mathcal D\leftarrow\emptyset$\;
 \While{training rollout budgets remain}{
  $\mathcal B_0\leftarrow\emptyset$; initialize $\mathbf s_1$ via~\eqref{eq:state}\;
  \For{$n=1,\ldots,M$}{
   $a_n\sim\pi_{\theta_{\mathrm{old}}}(\cdot|\mathbf s_n)$ masked by $\mathbf S_n^{\mathrm c}$\;
   $\mathcal B_n\leftarrow\mathcal B_{n-1}\cup\{z(a_n)\}$\; 
   Compute $R_n$ by~\eqref{eq:reward} and $\mathbf s_{n+1}$ via DT\;
   Set terminal indicator $d_n\leftarrow\mathbb I(n=M)$\;
   $\mathcal D\leftarrow\mathcal D\cup\{(\mathbf s_n,a_n,R_n,\mathbf s_{n+1},d_n)\}$\;
  }
  Update $\mathcal Q$ with $(\mathcal B_M,R(\mathcal B_M))$ via Section~\ref{buffer-section}\;
 }
 Compute advantages $\hat A_n$ and value targets from $\mathcal D$\;
 Update $\theta$ using~\eqref{eq:ppo_clip} and the mean squared error (MSE) value loss\;
}
\tcc{2. LS refinement on buffered deployments}
\For{$q=1,\ldots,|\mathcal Q|$}{
 \Repeat{no improvement in $R(\mathcal B^{(q)})$ over a full round}{
  \For{each $z_m\in\mathcal B^{(q)}$, $m=1,\ldots,M$}{
   $\Omega_m\leftarrow\{z\in\mathcal Z:\|z-z_m\|_2\le r_{\mathrm{ref}}\}$\;
   $z_m\leftarrow\arg\max_{\omega\in\Omega_m}R(\mathcal B^{(q)}_{z_m\leftarrow\omega})$ via DT\;
  }
 }
}
\KwRet{$\mathcal{B}^* \leftarrow \arg\max_{\mathcal{B} \in \mathcal{Q}} R(\mathcal{B})$\;}
\end{algorithm}

\section{Experimental Results}\label{sec:experiments}

This section evaluates both the geographic data-informed wireless network DT for radio map prediction and the DRL-based BS deployment optimization under realistic urban scenarios constructed from openly available geographic data.

\subsection{Experimental Settings}\label{subsec:exp_setting}

\subsubsection{Dataset and Scenario}
The experimental scenario is constructed using real-world geographic data from Xi'an, China. Building footprints and street networks are extracted from OpenStreetMap~\cite{haklay2008openstreetmap}. Following the procedure in Section~\ref{subsec:radiomap}, 80 urban areas are extracted and partitioned into 64 training, 8 validation, and 8 test areas. Each area is independently randomized 30 times with different material assignments based on Table~\ref{table:materials}. For each materialized area, 25 BS locations are randomly placed on building rooftops with antenna heights of 4~m above the rooftop, yielding total BS heights of 10--80~m. Receivers are placed at 1.5~m above the ground. Each BS transmits at 53~dBm RF output power at 3.5~GHz. The Sionna ray tracer is configured with diffraction enabled and a maximum reflection depth of 10. This yields $80 \times 30 \times 25 = 60{,}000$ radio map samples that jointly capture the effects of building geometry, height variation, BS placement, and material diversity.
For comparison, the USC dataset adopted in~\cite{lee_scalable_2024} covers less than 1~$\text{km}^2$ in total with uniform building heights and material properties, whereas our dataset spans over 60~$\text{km}^2$ with heterogeneous 3D building geometry, flexible BS heights, and diverse EM material assignments.

Real-world vehicle trajectory data recorded in Xi'an~\cite{didi2017gaia} are adopted for user mobility evaluation during rush hours. To ensure rigorous evaluation of Map2Traj, the training and test sets are geographically separated at longitude $108.974^\circ$. Map2Traj generates synthetic user trajectories conditioned on the underlying street network, providing user spatial distributions for the DT. Details about Map2Traj refer to~\cite{11299846}.
%,11299846 tao

For BS deployment, three scales with $M \in \{4, 5, 6\}$ BSs are evaluated on test areas that are unobserved during both radio map model and Map2Traj training, using eight test areas in Xi'an. The coverage threshold is set to $\psi_{\text{thr}} = -80$~dBm to ensure reliable connectivity in dense urban environments. Three coverage-capacity tradeoffs are evaluated with $\beta \in \{0.25, 0.5, 0.75\}$ in~\eqref{eq:objective_R}. The cross-city transfer experiment further uses real building data and DiDi user trajectories~\cite{didi2017gaia} from eight areas in Chengdu, a city unseen during DT training, within $30.63^\circ$--$30.65^\circ$N and $104.056^\circ$--$104.077^\circ$E. RT results based on building data and real trajectories provide ground truth for evaluating the final deployments.
\begin{table}[!t]
  \renewcommand{\arraystretch}{0.7}
  \centering
  \caption{Dataset and Scenario Settings}
  \label{table:dataset_config}
  \begin{tabular}{l c}
    \toprule
    \textbf{Parameter} & \textbf{Value} \\
    \midrule
    % \multicolumn{2}{c}{\textit{Dataset}} \\
    % \midrule
    Lat.\ / Lon.\ range (Xi'an) & $34.21^\circ$--$34.28^\circ$N \\
    & $108.912^\circ$--$108.996^\circ$E \\
    Total coverage & $\sim$~60 km$^2$ \\
    Area size & $1800 \times 1800$ m$^2$ \\
    &($256 \times 256$ pixels) \\
    Number of areas & 80 \\
    Total samples & 60{,}000 \\
    Carrier frequency & 3.5 GHz \\
    Ray tracing engine & Sionna~\cite{sionna} \\
    User trajectory data & DiDi (Xi'an, 2016)~\cite{didi2017gaia} \\
    Candidate BS locations $K$ & $\approx$10{,}000 per area \\
    Number of BSs $M$ & $\{4, 5, 6\}$ \\
    Coverage threshold $\psi_{\text{thr}}$ & $-80$ dBm \\
    Objective weight $\beta$ & $\{0.25, 0.5, 0.75\}$ \\
    \bottomrule
  \end{tabular}
\end{table}

\subsubsection{Model Settings and Training}
Three backbone architectures are evaluated for radio map prediction: a five-level U-Net~\cite{ronneberger2015u} with encoder-decoder structure and skip connections for preserving fine-grained spatial details; PMNet~\cite{lee2023pmnet}, which employs atrous spatial pyramid pooling (ASPP) for effective multi-scale feature extraction; and a ViT-based encoder-decoder model~\cite{11174709}. All radio map models are trained with the MSE loss and data augmentation technique described in Section~\ref{subsec:radiomap}. Learning rate scheduling (\textit{ReduceLROnPlateau}) and early stopping are employed to prevent overfitting.

For DRL-based BS deployment, the spatial PPO model described in Section~\ref{subsec:ppo} uses the five-channel state and $256\times256$ action grid. Each Xi'an configuration is trained for 1,000 episodes with 5 different seeds. A deployment buffer of capacity $N_{\text{buf}}=8$ is employed for post-training LS. Detailed model and training settings are summarized in Table~\ref{table:model_config}.
% PPO settings follow the top8_timed_formal campaign and DRL_exp_new/config.py.
\begin{table}[!t]
  \renewcommand{\arraystretch}{0.7}
  \centering
  \textcolor{black}{\caption{Model and Training Settings}}
  \label{table:model_config}
  \begin{tabular}{l c}
    \toprule
    \textbf{Parameter} & \textbf{Value} \\
    \midrule
    \multicolumn{2}{c}{\textit{Radio Map Prediction}} \\
    \midrule
    \textbf{U-Net} - Channel widths & $[64, 128, 256, 512, 1024]$ \\
    \textbf{PMNet} - Depths & $[3, 3, 27, 3]$ \\
    \textbf{PMNet} - Channel widths & $[256, 512, 512, 1024]$ \\
    \textbf{PMNet} - Atrous rates & $[6, 12, 18]$ \\
    \textbf{ViT} - Depths & $(2,6,14,2)$ \\
    \textbf{ViT} - Channel widths & $(128,256,512,1024)$ \\
    Learning rate / Weight decay & $10^{-4}$ / $10^{-5}$ \\
    Batch size & 32 \\
    \midrule
    \multicolumn{2}{c}{\textit{Spatial PPO}} \\
    \midrule
    Backbone (actor \& critic) & U-Net \\
    Training episodes $N_{\text{ep}}$ & 1{,}000 \\
    Learning rate & $3\times10^{-4}$ \\
    Discount factor $\gamma$, PPO clip $\epsilon$ & $1$ / $0.2$ \\
    Episodes per update, PPO epochs & 16, 4 \\
    Minibatch size, Entropy coefficient  & 4, 0.01 \\
    Deployment buffer capacity $N_{\text{buf}}$ & 8 \\
    Hyperparameters $\eta_{\max}, r_{\text{inh}}, r_{\text{ref}}$ & 10, 18, 25 \\
    \bottomrule
  \end{tabular}
\end{table}
\subsubsection{Compared Methods}
For radio map prediction, each backbone is evaluated under six input representations to quantify the contribution of each information modality:
\begin{itemize}
  \item \textbf{2D}: Binary building footprint map and binary BS map.
  \item \textbf{3D}: Building height map and BS map with antenna height encoding.
  \item \textbf{3D+EM}: Building height map, BS map, and EM material property maps, including conductivity and permittivity as described in Section~\ref{subsec:radiomap}.
  \item \textbf{X+Sample}: The above three configurations augmented with 100 sparse RSS samples provided as two additional input channels, including a sample mask and corresponding signal strengths. In this controlled comparison, these samples are sparse RT-generated RSS values from the same map.
\end{itemize}
Two metrics are adopted for radio map evaluation: MSE computed over outdoor pixels and coverage prediction accuracy (CPA), defined as the fraction of outdoor pixels where the predicted and ground truth coverage states agree given a signal strength threshold.

\begin{figure}[!t]
 \centering
 \includegraphics[width=\linewidth]{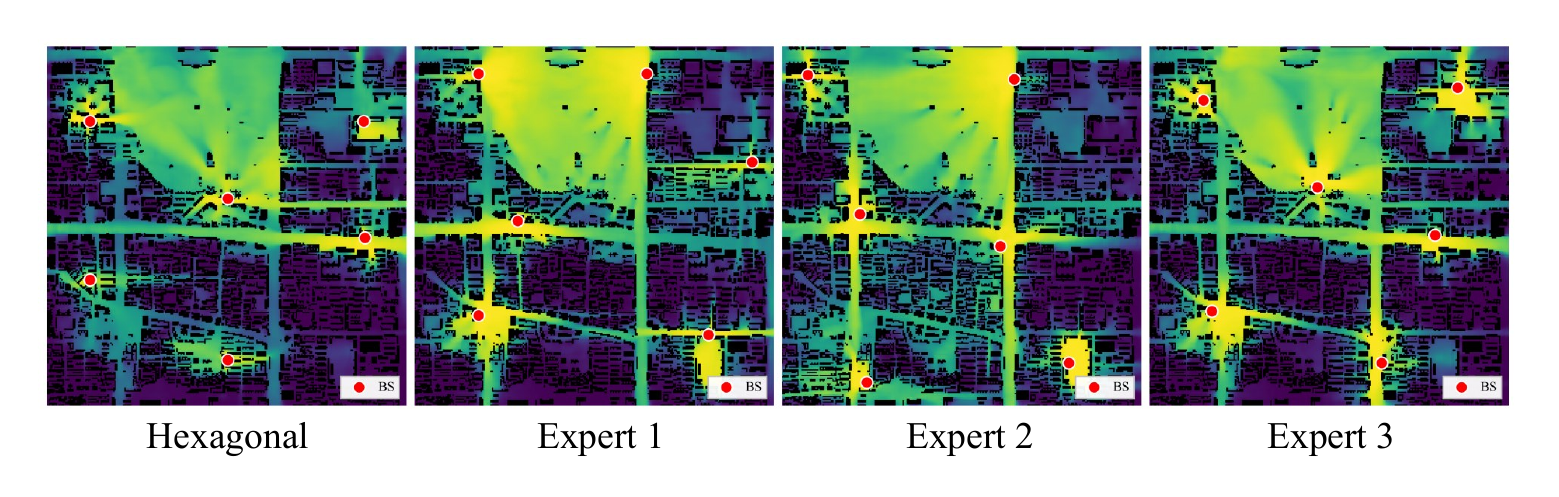}
 \caption{Geometric and three expert-designed deployments in Area~77.}
 \label{fig:expert}
\end{figure}
For BS deployment, seven methods are compared:
\begin{itemize}
 \item \textbf{Hexagonal}: BSs are arranged in regular hexagonal cells, a classical deployment pattern widely adopted in cellular network planning~\cite{borralho_survey_2021}.
 \item \textbf{Expert-mean}: The mean performance of three deployment configurations manually annotated by field experts based on city layouts, as illustrated in Fig.~\ref{fig:expert}.
 \item \textbf{Expert-max}: The best evaluated performance among the same three expert-designed configurations.
 \item \textbf{AutoBS~\cite{lee_autobs_2025}}: A state-of-the-art DT-based DRL method for BS deployment, using an MLP-based PPO agent and radio map prediction models. For a fair comparison in multi-BS deployment, its SNR-based capacity metric is replaced with the SINR-based counterpart in~\eqref{eq:sinr}.
 \item \textbf{AutoBS w/ our DT}: The AutoBS DRL model is retained, while reward calculation is replaced by~\eqref{eq:objective_R} via our DT.
 \item \textbf{Proposed method}: The DT in Section~\ref{sec:dt} is combined with the intelligent BS deployment framework in Section~\ref{sec:drl}, integrating spatial PPO exploration, Wasserstein distance-based deployment buffering, and LS refinement.
 \item \textbf{Genie-aided RT}: Using RT maps and real user trajectories, greedy selection evaluates all feasible sites at each step, followed by LS over an enlarged neighborhood.
\end{itemize}
These methods form three main categories. The first three are traditional methods whose deployment configurations require neither RT nor user distribution data. The next three are intelligent methods based on radio map prediction and DRL, with the latter two incorporating our DT to further account for user distribution. Genie-aided RT serves as a near-optimal reference based on the most extensive information and search. All resulting deployments are evaluated using the same RT maps, real user distribution, and weighted objective~\eqref{eq:objective_R}. The intelligent methods access these ground-truth data only during final evaluation, whereas Genie-aided RT uses them during optimization and Expert-max uses them to identify the best manual alternative.

\subsection{Radio Map Prediction Results}\label{subsec:exp_radiomap}

\subsubsection{Performance Comparison}

\begin{figure}[!t]
  \centering
  \subfloat[MSE comparison.\label{fig:mse_comparison}]{\includegraphics[width=0.9\linewidth]{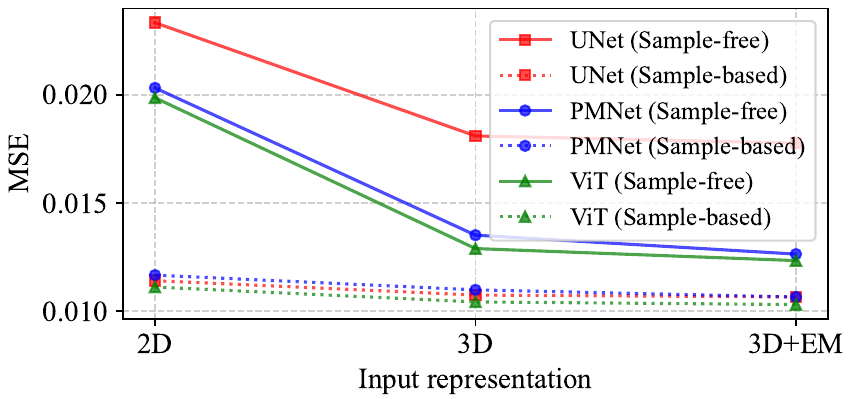}}
  \hfil
  \subfloat[CPA comparison.\label{fig:cpa_comparison}]{\includegraphics[width=0.9\linewidth]{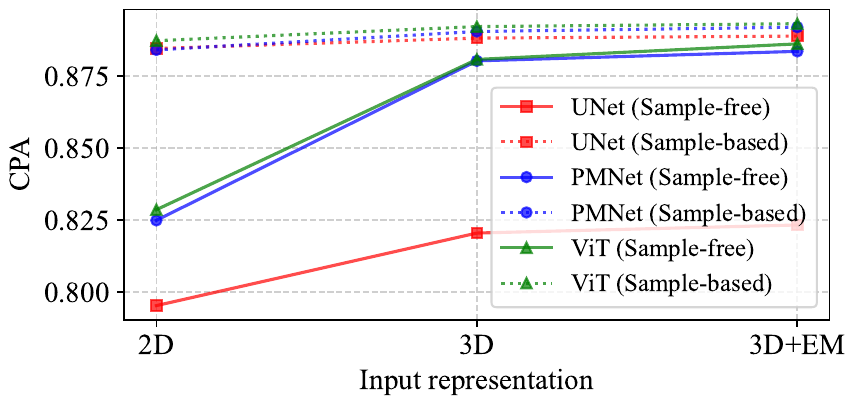}}
  \caption{Radio map prediction performance comparison.}
  \label{fig:radiomap_comparison}
\end{figure}

Fig.~\ref{fig:radiomap_comparison} summarizes the MSE and CPA results for U-Net, PMNet, and ViT with all six input representations. A consistent monotonic improvement is observed across all settings when sequentially augmenting the input from 2D to 3D and finally to 3D+EM. For instance, sample-free PMNet reduces its MSE from 0.0203 (2D) to 0.0135 (3D) and further to 0.0126 (3D+EM), corresponding to 33.5\% and 37.9\% reductions relative to the 2D baseline. This confirms that 3D building height information captures essential vertical propagation effects such as rooftop diffraction and line-of-sight blockage, while EM material properties further refine multipath modeling in reflection-dominant edge regions. As expected, incorporating 100 sparse RT-generated RSS samples improves prediction accuracy across all configurations. However, the performance gap narrows considerably with enriched input representations. For PMNet, the CPA difference between sample-free and sample-based settings decreases from 5.92 percentage points (2D) to merely 0.84 percentage points (3D+EM), with the sample-free 3D+EM setting achieving a CPA of 0.8836 compared to 0.8920 for its sample-based counterpart. 

It is also worth noting that the ability to exploit enriched input varies across architectures. Although the 3D and 3D+EM channels improve U-Net, a substantial gap to its sample-based counterpart persists, indicating that U-Net's limited receptive field constrains its capacity to utilize the additional geometric and material information. In contrast, PMNet and ViT, which leverage multi-scale or global attention mechanisms, effectively exploit the enriched 3D+EM representation and nearly close the gap to their sample-based counterparts in terms of both MSE and CPA.

These results collectively validate that the proposed 3D+EM hybrid input representation captures the dominant propagation mechanisms in urban macro BS scenarios, enabling near-sample-based prediction accuracy in a purely geographic data-driven, sample-free manner. We adopt PMNet for subsequent radio map analysis and BS deployment optimization, as it offers a favorable tradeoff between prediction accuracy and computational efficiency.

\subsubsection{Qualitative Analysis}

% \begin{figure*}[!t]
%   \centering
%   \subfloat[Building map]{\includegraphics[width=0.18\linewidth]{Fig/Radiomap/4303_building.pdf}}
%   \hfil
%   \subfloat[Ray tracing result]{\includegraphics[width=0.18\linewidth]{Fig/Radiomap/4303_gt.pdf}}
%   \hfil
%   \subfloat[2D]{\includegraphics[width=0.18\linewidth]{Fig/Radiomap/4303_pmnet_2d.pdf}}
%   \hfil
%   \subfloat[3D]{\includegraphics[width=0.18\linewidth]{Fig/Radiomap/4303_pmnet_3d.pdf}}
%   \hfil
%   \subfloat[3D+EM]{\includegraphics[width=0.18\linewidth]{Fig/Radiomap/4303_pmnet_3d_em.pdf}}

%   \subfloat[Building map]{\includegraphics[width=0.18\linewidth]{Fig/Radiomap/5637_building.pdf}}
%   \hfil
%   \subfloat[Ray tracing result]{\includegraphics[width=0.18\linewidth]{Fig/Radiomap/5637_gt.pdf}}
%   \hfil
%   \subfloat[2D]{\includegraphics[width=0.18\linewidth]{Fig/Radiomap/5637_pmnet_2d.pdf}}
%   \hfil
%   \subfloat[3D]{\includegraphics[width=0.18\linewidth]{Fig/Radiomap/5637_pmnet_3d.pdf}}
%   \hfil
%   \subfloat[3D+EM]{\includegraphics[width=0.18\linewidth]{Fig/Radiomap/5637_pmnet_3d_em.pdf}}
%   \caption{Qualitative comparison of PMNet radio map predictions under 2D, 3D, and 3D+EM inputs.}
%   \label{fig:radiomap}
% \end{figure*}
\begin{figure}[!t]
  \centering
  \subfloat[RT]{\includegraphics[width=0.25\linewidth,trim=15 12 15 12, clip]{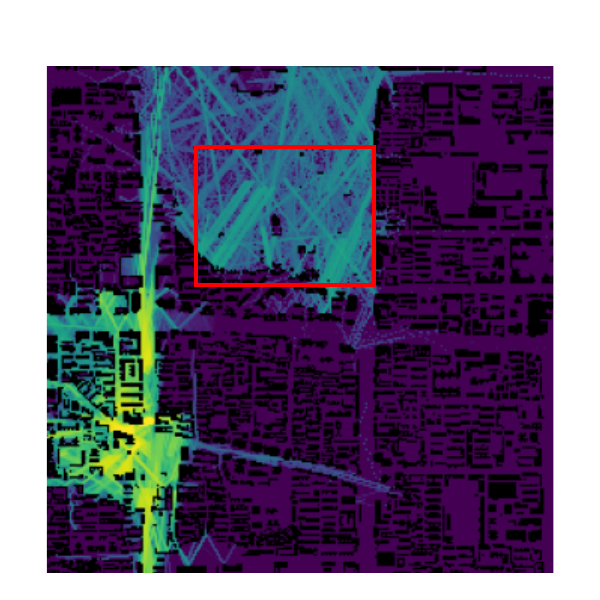}}
  \hfil
  \subfloat[2D]{\includegraphics[width=0.25\linewidth,trim=15 12 15 12, clip]{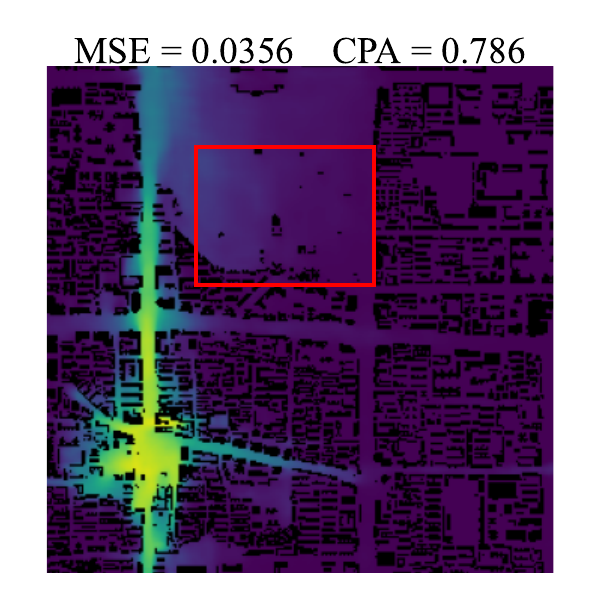}}
  \hfil
  \subfloat[3D]{\includegraphics[width=0.25\linewidth,trim=15 12 15 12, clip]{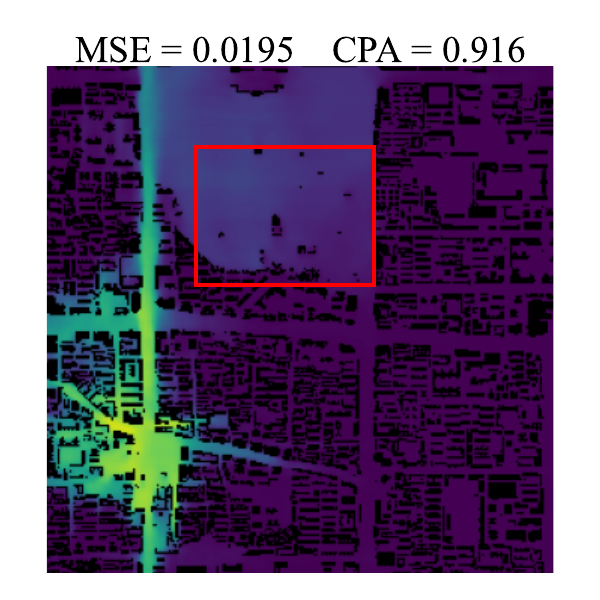}}
  \hfil
  \subfloat[3D+EM]{\includegraphics[width=0.25\linewidth,trim=15 12 15 12, clip]{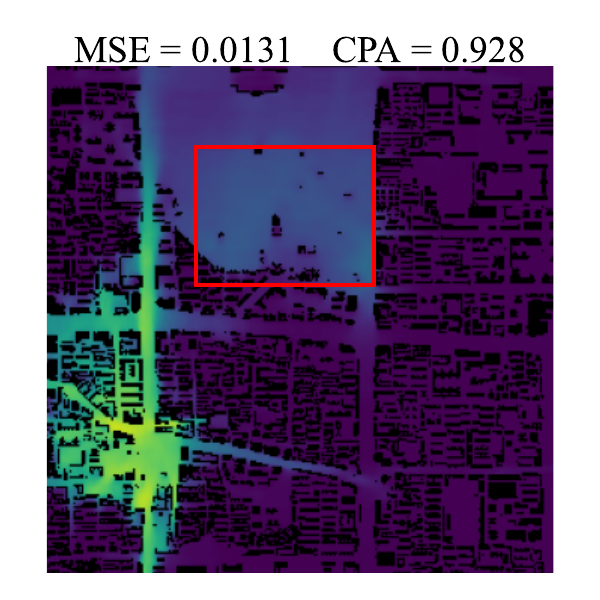}}

  \subfloat[RT]{\includegraphics[width=0.25\linewidth,trim=15 12 15 12, clip]{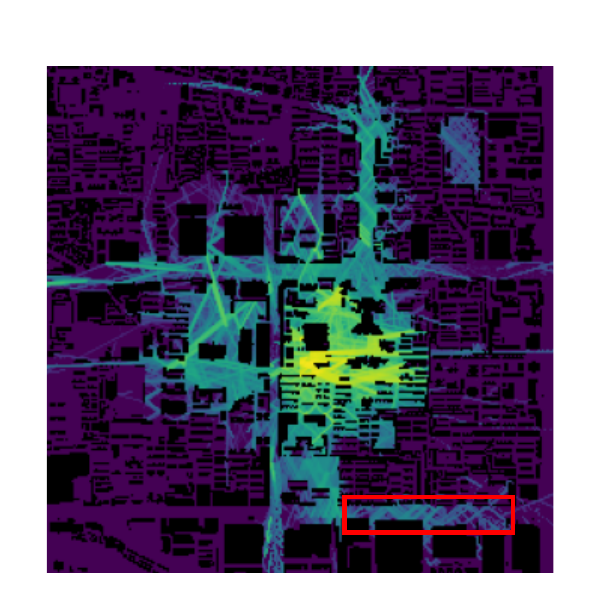}}
  \hfil
  \subfloat[2D]{\includegraphics[width=0.25\linewidth,trim=15 12 15 12, clip]{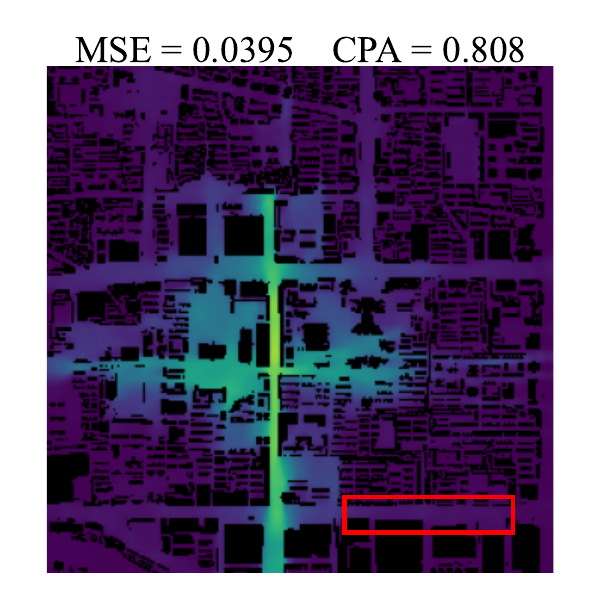}}
  \hfil
  \subfloat[3D]{\includegraphics[width=0.25\linewidth,trim=15 12 15 12, clip]{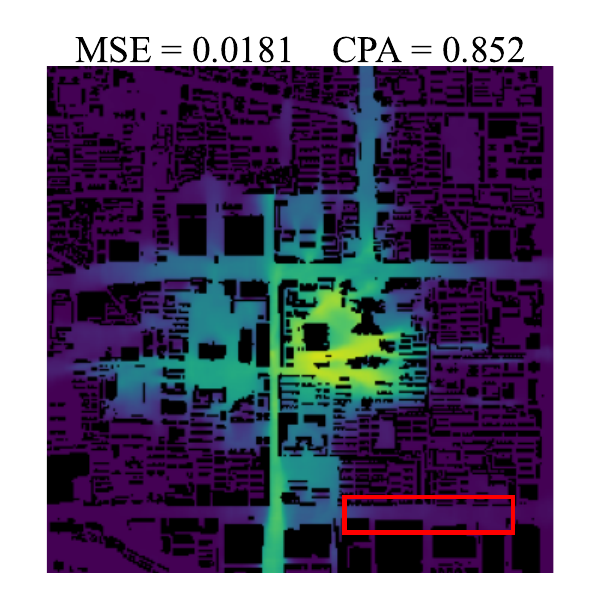}}
  \hfil
  \subfloat[3D+EM]{\includegraphics[width=0.25\linewidth,trim=15 12 15 12, clip]{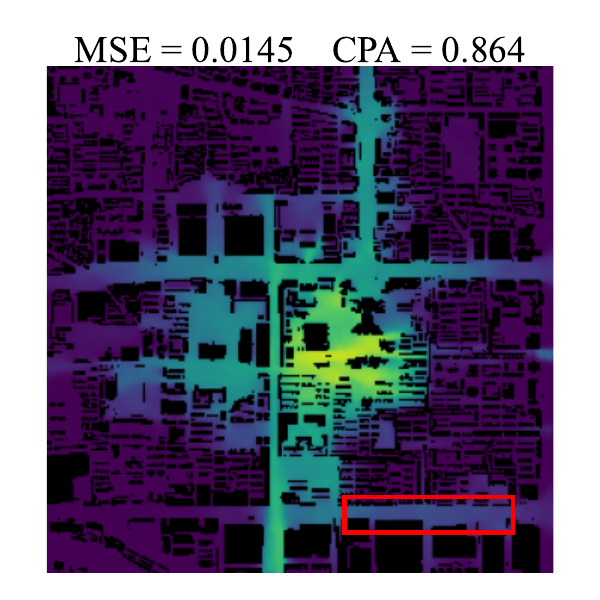}}
  \caption{Qualitative comparison of PMNet radio map predictions (upper: Area~77; lower: Area~79).}
  \label{fig:radiomap}
\end{figure}

Fig.~\ref{fig:radiomap} provides a qualitative comparison of PMNet predictions under different input representations across two representative test areas. The 2D model produces overly conservative predictions and fails to recover signal coverage in unobstructed regions, since it cannot distinguish low-rise buildings that permit over-rooftop propagation from tall structures that fully block the signal. By incorporating building height information, the 3D model captures vertical obstruction and over-rooftop diffraction, substantially improving prediction accuracy in the vicinity of the BS. However, the 3D input still underestimates signal strength in reflection-dominated peripheral areas, as highlighted by the red boxes in Fig.~\ref{fig:radiomap}. Augmenting the input with EM material properties refines multipath modeling and corrects these boundary-region underestimations. Quantitatively, this progression yields a 63.2\% MSE reduction from 0.0356 (2D) to 0.0131 (3D+EM) in the upper area, and a 63.3\% reduction to 0.0145 in the lower area. These results confirm that 3D geometry captures height-dependent propagation effects, while EM properties further enhance accuracy in reflection-prone edge regions, jointly validating the effectiveness of the proposed hybrid input representation for radio map prediction in urban macro BS scenarios.

\subsection{BS Deployment Optimization Results}\label{subsec:exp_deploy}
Building upon the validated wireless network DT with PMNet (3D+EM), we evaluate the intelligent BS deployment framework from four aspects: the evolution of BS deployments during DRL training, optimization performance in held-out Xi'an test areas, component-wise ablation of the intelligent deployment algorithm, and cross-city transfer to Chengdu. Final deployments from all methods are assessed using ground-truth RT radio maps and real user trajectories.

\subsubsection{Deployment Evolution During Training}
\begin{figure}[!t]
\centering
\includegraphics[width=1\linewidth]{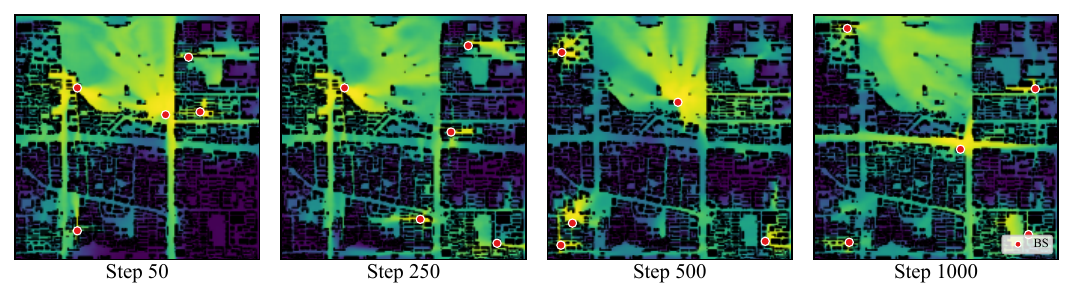}
\caption{Evolution of deployment during training in Area~77.}
\label{fig:training_dynamics}
\end{figure}

Fig.~\ref{fig:training_dynamics} illustrates the evolution of BS deployments during DRL training for $M=5$ and $\beta=0.5$ in a representative test area. At episode 100, the deployment contains spatially clustered BSs with substantial inter-BS interference and coverage overlap, leaving substantial portions of the area underserved. As training progresses, the selected BSs become more dispersed and extend service to previously uncovered regions. The final configuration exhibits broader spatial coverage and less redundant overlap, demonstrating the ability to balance coverage expansion, interference, and capacity under the objective.

\subsubsection{Deployment Performance in Held-Out Xi'an Areas}
Table~\ref{table:deploy_main} compares the seven methods over held-out Xi'an test areas, with $M\in\{4,5,6\}$ and $\beta\in\{0.25,0.5,0.75\}$. The reported optimization time includes DRL training, DT inference, LS post-processing, and buffer storage and replacement operations, and excludes final ground-truth evaluation. Since the objective uses the same estimated user distribution $\hat{\rho}$ across BS locations, BS counts, and optimization methods for a given area, map-conditioned trajectory generation and user distribution estimation are performed offline before optimization and excluded from the reported time. For Genie-aided RT, the reported time is predominantly determined by the RT over all feasible BS locations, resulting in a consistent $31.5$~h across all settings.

\begin{table*}[!t]
\centering
\footnotesize
\setlength{\tabcolsep}{2pt}
\renewcommand{\arraystretch}{0.9}
\caption{BS Deployment Results in Xi'an}
\label{table:deploy_main}
\begin{threeparttable}
\begin{tabular}{cll*{9}{c}c}
\toprule
\multirow{2}{*}{$M$} & \multirow{2}{*}{Category} & \multirow{2}{*}{Method} & \multicolumn{3}{c}{$\beta=0.25$} & \multicolumn{3}{c}{$\beta=0.5$} & \multicolumn{3}{c}{$\beta=0.75$} & \multirow{2}{*}{\shortstack{Optimization\\time}} \\
\cmidrule(lr){4-6}\cmidrule(lr){7-9}\cmidrule(lr){10-12}
& & & Objective & Coverage & Capacity & Objective & Coverage & Capacity & Objective & Coverage & Capacity & \\
\midrule
\multirow{7}{*}{4} & \multirow{3}{*}{Traditional} & Hexagonal & 0.497 & 64.0\% & 0.449 & 0.545 & 64.0\% & 0.449 & 0.592 & 64.0\% & 0.449 & \textemdash \\
 & & Expert-mean & 0.543 & 66.6\% & 0.502 & 0.584 & 66.6\% & 0.502 & 0.625 & 66.6\% & 0.502 & \textemdash \\
 & & Expert-max & 0.577 & 69.1\% & 0.539 & 0.615 & 69.1\% & 0.539 & 0.660 & 71.4\% & 0.497 & \textemdash \\
\addlinespace[1.5pt]
 & \multirow{3}{*}{Intelligent} & AutoBS~\cite{lee_autobs_2025} & 0.534 & 68.0\% & 0.485 & 0.590 & 70.6\% & 0.475 & 0.676 & 74.2\% & 0.477 & $117.1$ s \\
 & & AutoBS w/ our DT & 0.572 & 65.1\% & 0.546 & 0.621 & 72.0\% & 0.522 & 0.679 & 74.3\% & 0.484 & $117.1$ s \\
 & & Proposed method & \textbf{0.672} & \textbf{76.1\%} & \textbf{0.642} & \textbf{0.717} & \textbf{81.2\%} & \textbf{0.623} & \textbf{0.772} & \textbf{83.6\%} & \textbf{0.577} & $184.4$ s \\
\addlinespace[1.5pt]
 & Near-optimal & Genie-aided RT & 0.722 & 75.5\% & 0.710 & 0.747 & 79.9\% & 0.696 & 0.796 & 85.6\% & 0.616 & $31.5$ h \\
\midrule
\multirow{7}{*}{5} & \multirow{3}{*}{Traditional} & Hexagonal & 0.519 & 69.7\% & 0.460 & 0.579 & 69.7\% & 0.460 & 0.638 & 69.7\% & 0.460 & \textemdash \\
 & & Expert-mean & 0.623 & 70.0\% & 0.597 & 0.649 & 70.0\% & 0.597 & 0.674 & 70.0\% & 0.597 & \textemdash \\
 & & Expert-max & 0.659 & 69.0\% & 0.649 & 0.675 & 72.7\% & 0.624 & 0.705 & 74.1\% & 0.599 & \textemdash \\
\addlinespace[1.5pt]
 & \multirow{3}{*}{Intelligent} & AutoBS & 0.617 & 72.0\% & 0.583 & 0.648 & 73.8\% & 0.558 & 0.714 & 77.5\% & 0.533 & $142.0$ s \\
 & & AutoBS w/ our DT & 0.652 & 68.8\% & 0.640 & 0.676 & 73.7\% & 0.615 & 0.722 & 77.4\% & 0.567 & $142.7$ s \\
 & & Proposed method & \textbf{0.766} & \textbf{77.6\%} & \textbf{0.763} & \textbf{0.786} & \textbf{82.3\%} & \textbf{0.749} & \textbf{0.815} & \textbf{85.7\%} & \textbf{0.688} & $232.3$ s \\
\addlinespace[1.5pt]
 & Near-optimal & Genie-aided RT & 0.849 & 78.0\% & 0.872 & 0.838 & 81.4\% & 0.861 & 0.845 & 87.7\% & 0.750 & $31.5$ h \\
\midrule
\multirow{7}{*}{6} & \multirow{3}{*}{Traditional} & Hexagonal & 0.627 & 73.3\% & 0.591 & 0.662 & 73.3\% & 0.591 & 0.697 & 73.3\% & 0.591 & \textemdash \\
 & & Expert-mean & 0.687 & 73.8\% & 0.670 & 0.704 & 73.8\% & 0.670 & 0.721 & 73.8\% & 0.670 & \textemdash \\
 & & Expert-max & 0.733 & 72.3\% & 0.736 & 0.730 & 72.3\% & 0.736 & 0.737 & 75.6\% & 0.682 & \textemdash \\
\addlinespace[1.5pt]
 & \multirow{3}{*}{Intelligent} & AutoBS & 0.677 & 74.9\% & 0.652 & 0.702 & 77.0\% & 0.635 & 0.755 & 80.7\% & 0.597 & $165.7$ s \\
 & & AutoBS w/ our DT & 0.718 & 73.0\% & 0.714 & 0.716 & 75.7\% & 0.676 & 0.756 & 79.2\% & 0.648 & $166.9$ s \\
 & & Proposed method & \textbf{0.859} & \textbf{78.6\%} & \textbf{0.883} & \textbf{0.845} & \textbf{83.5\%} & \textbf{0.855} & \textbf{0.856} & \textbf{86.9\%} & \textbf{0.814} & $277.9$ s \\
\addlinespace[1.5pt]
 & Near-optimal & Genie-aided RT & 0.963 & 81.0\% & 1.014 & 0.915 & 83.0\% & 1.000 & 0.888 & 88.7\% & 0.891 & $31.5$ h \\
\bottomrule
\end{tabular}
\begin{tablenotes}\footnotesize
\item Results are averaged over eight areas with five PPO seeds per area. Bold indicates the best result excluding Genie-aided RT.
\end{tablenotes}
\end{threeparttable}
\end{table*}

The proposed method consistently achieves substantially higher objectives than hexagonal and expert-designed deployments across all nine configurations. AutoBS outperforms hexagonal placement throughout these settings, but falls below Expert-max in six of the nine configurations. As discussed in Section~\ref{subsec:capacity}, the evaluated AutoBS baseline lacks user distribution information and estimates capacity from predicted radio-map SINR without accounting for user-dependent resource sharing. Our DT additionally predicts user trajectories from geographic data, enabling capacity estimation that accounts for spatial user demand and resource sharing. By incorporating our DT into the AutoBS framework, we achieve consistent improvements in all nine settings. Building on this DT, the proposed intelligent deployment algorithm provides a further improvement through spatial policy learning and deployment refinement.

For example, with $M=5$ and $\beta=0.5$, the objective increases from 0.648 for AutoBS to 0.676 with our DT and further to 0.786 with the complete proposed method, compared with 0.579 for Hexagonal and 0.675 for Expert-max. The proposed method also adapts its optimization direction to the selected coverage-capacity tradeoff: smaller $\beta$ prioritizes capacity, whereas larger $\beta$ places greater emphasis on coverage. For $M=5$, increasing $\beta$ from 0.25 to 0.75 raises coverage from 77.6\% to 85.7\%, while normalized capacity decreases from 0.763 to 0.688. This adjustment demonstrates that the method can accommodate different deployment preferences through the objective weight.

Across the evaluated settings, the proposed method attains approximately 89.2\%--97.0\% of the objective achieved by the near-optimal Genie-aided RT reference. Meanwhile, optimization time decreases from approximately 31.5 h for Genie-aided RT to roughly four minutes per area. These results demonstrate a favorable balance between deployment quality and computational cost without requiring ground-truth radio maps or real user trajectories during optimization.

\subsubsection{Ablation of the Intelligent Deployment Algorithm}\label{subsec:ablation}
Table~\ref{table:deploy_ablation} presents a cumulative ablation at a representative setting, i.e., $\beta=0.5$ and $M=5$, over eight Xi'an areas with five seeds. All variants use the proposed DT and the same five-channel state representation, and the compared DRL models have similar parameter counts for a fair architectural comparison. The ablation therefore isolates the contributions of the intelligent deployment algorithm from DT design. Compared with the MLP-based AutoBS model~\cite{lee_autobs_2025}, the spatial PPO model increases the objective from 0.676 to 0.732, while reducing runtime from 144.1 to 117.1 s. These results support the advantage of spatial feature extraction for processing kilometer-scale urban states, improving both deployment performance and computational efficiency.

\begin{table}[!t]
\centering
\footnotesize
\caption{Deployment Ablation for Intelligent Deployment Algorithm}
\label{table:deploy_ablation}
\setlength{\tabcolsep}{5.5pt}
\begin{threeparttable}
\begin{tabular}{lcccc}
\toprule
Method & Objective & Coverage & Capacity & Time (s) \\
\midrule
AutoBS model~\cite{lee_autobs_2025} & 0.676 & 73.7\% & 0.615 & $144.1$ \\
Spatial PPO & 0.732 & 77.4\% & 0.689 & $117.1$ \\
Spatial PPO+LS & 0.768 & 80.0\% & 0.736 & $144.0$ \\
Spatial PPO+buffer+LS & \textbf{0.786} & \textbf{82.3\%} & \textbf{0.749} & $231.9$ \\
\bottomrule
\end{tabular}
\begin{tablenotes}\footnotesize
\item All variants are trained within our DT.
\end{tablenotes}
\end{threeparttable}
\end{table}

Post-processing provides further gains beyond spatial policy learning. Adding LS raises the objective to 0.768, and applying LS to the eight buffered deployments further increases it to 0.786. The corresponding runtimes increase to 144.0 and 231.9 s, respectively, with the additional cost arising mainly from DT inferences to evaluate radio maps at neighboring candidate BS locations during LS. In summary, local refinement improves the learned deployments, while the buffer supplies multiple starting configurations that enable further improvement.

\subsubsection{Cross-City Transfer}\label{subsec:transfer}
\begin{figure}[!t]
 \centering
 \includegraphics[width=1\linewidth]{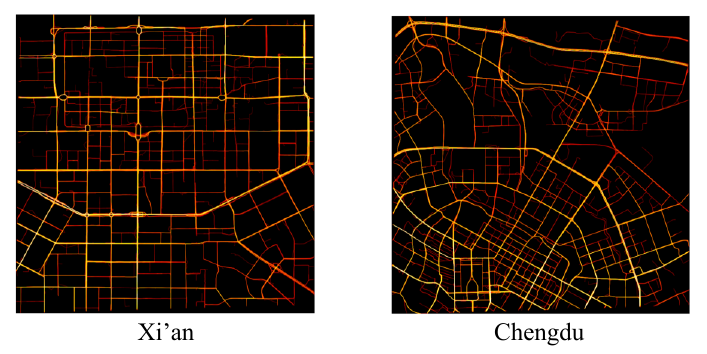}
 \caption{City-level trajectory heatmaps for Xi'an and Chengdu.}
 \label{fig:city_layout}
\end{figure}

Xi'an and Chengdu are two cities in China that exhibit pronounced differences in urban morphology and user mobility patterns, as illustrated in Fig.~\ref{fig:city_layout}, providing a test of transfer across distinct urban environments. Both the radio map predictor and user trajectory generation model trained on Xi'an are applied directly without fine-tuning on target-city data. Using these trained models, we construct the DT for each Chengdu area solely from its 3D building data and street maps and run the proposed intelligent BS deployment algorithm within this DT. Target-city RT maps and real user trajectories are used only for final evaluation.

\begin{table*}[!t]
\centering
\footnotesize
\setlength{\tabcolsep}{3pt}
\renewcommand{\arraystretch}{0.9}
\caption{Cross-City Deployment Results in Chengdu}
\label{table:deploy_transfer}
\begin{threeparttable}
\begin{tabular}{cll*{9}{c}c}
\toprule
\multirow{2}{*}{$M$} & \multirow{2}{*}{Category} & \multirow{2}{*}{Method} & \multicolumn{3}{c}{$\beta=0.25$} & \multicolumn{3}{c}{$\beta=0.5$} & \multicolumn{3}{c}{$\beta=0.75$} & \multirow{2}{*}{\shortstack{Optimization\\time}} \\
\cmidrule(lr){4-6}\cmidrule(lr){7-9}\cmidrule(lr){10-12}
& & & Objective & Coverage & Capacity & Objective & Coverage & Capacity & Objective & Coverage & Capacity & \\
\midrule
\multirow{7}{*}{4} & \multirow{3}{*}{Traditional} & Hexagonal & 0.614 & 73.8\% & 0.573 & 0.655 & 73.8\% & 0.573 & 0.696 & 73.8\% & \textbf{0.573} & \textemdash \\
 & & Expert-mean & 0.614 & 80.5\% & 0.550 & 0.678 & 80.5\% & 0.550 & 0.741 & 80.5\% & 0.550 & \textemdash \\
 & & Expert-max & 0.642 & 79.6\% & 0.591 & 0.696 & 81.9\% & 0.573 & 0.762 & 83.2\% & 0.553 & \textemdash \\
\addlinespace[1.5pt]
 & \multirow{3}{*}{Intelligent} & AutoBS & 0.599 & 80.2\% & 0.531 & 0.671 & 82.2\% & 0.519 & 0.751 & 84.3\% & 0.477 & $106.0$ s \\
 & & AutoBS w/ our DT & 0.630 & 77.4\% & 0.582 & 0.683 & 82.2\% & 0.545 & 0.769 & 86.4\% & 0.486 & $106.0$ s \\
 & & Proposed method & \textbf{0.684} & \textbf{83.6\%} & \textbf{0.633} & \textbf{0.751} & \textbf{88.3\%} & \textbf{0.619} & \textbf{0.829} & \textbf{92.9\%} & 0.530 & $128.0$ s \\
\addlinespace[1.5pt]
 & Near-optimal & Genie-aided RT & 0.737 & 83.7\% & 0.704 & 0.776 & 87.6\% & 0.676 & 0.841 & 92.9\% & 0.575 & $15.2$ h \\
\midrule
\multirow{7}{*}{5} & \multirow{3}{*}{Traditional} & Hexagonal & 0.669 & 75.6\% & 0.640 & 0.698 & 75.6\% & 0.640 & 0.727 & 75.6\% & 0.640 & \textemdash \\
 & & Expert-mean & 0.678 & 86.0\% & 0.618 & 0.739 & 86.0\% & 0.618 & 0.799 & 86.0\% & 0.618 & \textemdash \\
 & & Expert-max & 0.721 & \textbf{88.4\%} & 0.667 & 0.776 & 88.4\% & 0.667 & 0.830 & 88.4\% & \textbf{0.667} & \textemdash \\
\addlinespace[1.5pt]
 & \multirow{3}{*}{Intelligent} & AutoBS & 0.689 & 81.6\% & 0.647 & 0.725 & 85.1\% & 0.600 & 0.797 & 87.6\% & 0.562 & $126.4$ s \\
 & & AutoBS w/ our DT & 0.716 & 81.2\% & 0.684 & 0.748 & 83.5\% & 0.662 & 0.801 & 87.1\% & 0.591 & $126.8$ s \\
 & & Proposed method & \textbf{0.782} & 85.4\% & \textbf{0.758} & \textbf{0.817} & \textbf{90.6\%} & \textbf{0.728} & \textbf{0.864} & \textbf{94.2\%} & 0.629 & $154.0$ s \\
\addlinespace[1.5pt]
 & Near-optimal & Genie-aided RT & 0.857 & 85.9\% & 0.857 & 0.864 & 88.9\% & 0.838 & 0.889 & 93.7\% & 0.745 & $15.2$ h \\
\midrule
\multirow{7}{*}{6} & \multirow{3}{*}{Traditional} & Hexagonal & 0.763 & 79.4\% & 0.753 & 0.774 & 79.4\% & 0.753 & 0.784 & 79.4\% & 0.753 & \textemdash \\
 & & Expert-mean & 0.751 & 89.2\% & 0.703 & 0.798 & 89.2\% & 0.703 & 0.845 & 89.2\% & 0.703 & \textemdash \\
 & & Expert-max & 0.810 & \textbf{89.4\%} & 0.782 & 0.840 & 90.2\% & 0.779 & 0.872 & 90.4\% & \textbf{0.776} & \textemdash \\
\addlinespace[1.5pt]
 & \multirow{3}{*}{Intelligent} & AutoBS & 0.747 & 83.2\% & 0.719 & 0.796 & 88.0\% & 0.712 & 0.845 & 89.8\% & 0.684 & $145.9$ s \\
 & & AutoBS w/ our DT & 0.779 & 82.8\% & 0.763 & 0.802 & 85.5\% & 0.750 & 0.838 & 89.8\% & 0.659 & $146.2$ s \\
 & & Proposed method & \textbf{0.874} & 88.5\% & \textbf{0.870} & \textbf{0.881} & \textbf{90.7\%} & \textbf{0.854} & \textbf{0.900} & \textbf{95.1\%} & 0.749 & $176.9$ s \\
\addlinespace[1.5pt]
 & Near-optimal & Genie-aided RT & 1.000 & 85.1\% & 1.049 & 0.957 & 90.1\% & 1.014 & 0.937 & 93.5\% & 0.943 & $15.2$ h \\
\bottomrule
\end{tabular}
\end{threeparttable}
\end{table*}

Table~\ref{table:deploy_transfer} shows that the proposed method retains the highest objective among the traditional and intelligent deployment methods across all nine Chengdu settings, attaining approximately 87.4\%--98.6\% of the near-optimal Genie-aided RT objective.

Compared with Xi'an, where building density is relatively high, Chengdu's lower building density makes expert-selected BS locations more competitive, allowing expert deployments to outperform the proposed method on individual metrics in some settings. However, the proposed method consistently achieves the highest joint optimization objective among the traditional and intelligent deployment methods. For example, with $M=5$ and $\beta=0.25$, Expert-max provides slightly higher coverage, but the proposed method achieves a higher objective, 0.782 versus 0.721, through improved capacity. The lower building density also leads to a substantially reduced number of feasible BS deployment locations, approximately 6{,}000 per area, requiring fewer DT evaluations during post-processing. Consequently, the proposed method completes deployment optimization in less than three minutes per area. Overall, these results demonstrate the cross-city generalization capability of the proposed framework, which uses only geographic data from the target city to achieve deployment performance approaching the Genie-aided RT reference while substantially reducing optimization time.

\section{Conclusion}\label{sec:conclusion}
In this paper, we have proposed a geographic data-informed wireless network DT framework for intelligent urban macro BS deployment. By combining radio map prediction with map-conditioned trajectory generation, the DT enables coverage-capacity evaluation that jointly accounts for propagation, inter-BS interference, and spatial user demand through user association and bandwidth sharing. The DT framework supports deployment optimization without target-area measurements, RT, or real user trajectories. Building upon this DT, we have formulated the BS deployment process as a multi-step MDP and developed a spatial PPO model enhanced with LS refinement and a Wasserstein distance-based deployment buffer, which combines policy exploration with solution-level refinement in the combinatorial deployment space. Experiments demonstrate that the hybrid representation achieves sample-free radio map accuracy close to prediction using 100 sparse RSS samples. Deployment results in held-out Xi'an areas show consistent objective improvements over traditional and DRL baselines, while component-wise ablations confirm the benefits of spatial policy learning and subsequent refinement. The complete framework approaches the near-optimal solution while reducing optimization time from tens of hours to approximately four minutes per area. Furthermore, direct transfer to Chengdu without DT model fine-tuning achieves 87.4\%--98.6\% of the near-optimal performance in less than three minutes per area. These results demonstrate the effectiveness, efficiency, and cross-city generalization of the framework, supporting high-quality BS deployment optimization using only geographic data from the target area.

\footnotesize
\bibliographystyle{IEEEtran}
\bibliography{IEEEabrv,IEEEexample}
\end{document}